\documentclass[prb,twocolumn,floatfix]{revtex4-1}

\usepackage{amsfonts}
\usepackage{amsmath}
\usepackage{verbatim}
\usepackage[dvips]{graphicx}
\usepackage{subfigure}

\def\Tr{\textrm}
\def\dd{\textrm{d}}
\def\Bf{\boldsymbol}
\def\Wt{\widetilde}

\def\rv{\Bf{r}}
\def\Rv{\Bf{R}}

\def\kv{\Bf{k}}
\def\pv{\Bf{p}}
\def\qv{\Bf{q}}

\def\Gv{\Bf{G}}
\def\nv{\Bf{n}}
\def\mv{\Bf{m}}

\def\EE{\varepsilon}
\def\EB{\epsilon}

\begin{document}

\title{$s$-wave Cooper pair insulators and theory of correlated superconductors}
\author{Predrag Nikoli\'{c}$^{1,2}$ and Zlatko Te\v{s}anovi\'{c}$^{2}$}
\affiliation{$^1$Department of Physics and Astronomy, George Mason University, Fairfax, VA 22030, USA}
\affiliation{$^2$Institute for Quantum Matter and Department of Physics and Astronomy, Johns Hopkins University, Baltimore, MD 21218, USA}
\date{\today}

\begin{abstract}

The pseudogap state of cuprate high-temperature superconductors has been often viewed as either a yet unknown competing order or a precursor state to superconductivity. While awaiting the resolution of the pseudogap problem in cuprates, we demonstrate that local pairing fluctuations, vortex liquid dynamics and other precursor phenomena can emerge quite generally whenever fermionic excitations remain gapped across the superconducting transition, regardless of the gap origin. Our choice of a tractable model is a lattice band insulator with short-range attractive interactions between fermions in the $s$-wave channel. An effective crossover between Bardeen-Cooper-Schrieffer (BCS) and Bose-Einstein condensate (BEC) regimes can be identified in any band insulator above two dimensions, while in two dimensions only the BEC regime exists. The superconducting transition is ``unconventional'' (non-pair-breaking) in the BEC regime, identified by either the bosonic mean-field or XY universality class. The insulator adjacent to the superconductor in the BEC regime is a bosonic Mott insulator of Cooper pairs, which may be susceptible to charge density wave ordering. We construct a function of the many-body excitation spectrum whose non-analytic changes define a sharp distinction between band and Mott insulators. The corresponding ``second order transition'' can be observed out of equilibrium by driving a Cooper pair laser in the Mott insulator. We explicitly show that the gap for charged bosonic excitations lies below the threshold for Cooper pair breakup in any BEC regime, despite quantum fluctuations. Our discussion ends with a view of possible consequences for cuprates, where antinodal pair dynamics has certain features in common with our simple $s$-wave picture.

\end{abstract}

\maketitle

\section{Introduction}

The microscopic origin of high temperature superconductivity, particularly
in copper oxides, remains one of the greatest unresolved
challenges in condensed matter physics. All systems known to superconduct at relatively high temperatures exhibit strong electronic correlations and great complexity, which the standard well-controlled theoretical techniques are ill-equipped to handle. Most novel approaches to correlated superconductors either focus on spin dynamics \cite{anderson87c, Wen1995, senthil00, Chakravarty2001, anderson04, Kaul2007}, or charge dynamics \cite{Doniach1990, emery95, balents98, franz02, tesanovic04, balents05} in their attempt to explain the multitude of unconventional properties of the correlated normal states. This is naturally
justified by proximity to either antiferromagnetic or superconducting phases.

The charge-oriented approaches are based on the assumption that strong pairing correlations, or locally stable Cooper pairs, can be held responsible for the pseudogap and normal state features. This ``precursor'' picture indeed finds support in a number of experiments \cite{Ando1995, Corson1999, Fang2004, Yazdani2004, Wang2005, Ong2007, Armitage2007}. If so, a prominent physical description of such normal states, originating in duality \cite{Fisher1989}, is a liquid of quantum vortices and antivortices,
whose free motion dynamically disorders the phase of the complex order parameter and
thus preemts "Bose condensation" of such locally stable Cooper pairs.

The freedom in principle to either form or not form locally stable Cooper pairs opens a possibility of having two kinds of normal states. Conventional normal states have unpaired fermions, and are exemplified by Fermi liquids, band insulators and integer quantum Hall states (IQHS). The second kind are unconventional normal states which involve manifestations of Cooper pair stability and exhibit strong correlations. Examples of such states discussed in literature are pair density waves (PDW) \cite{tesanovic04, balents05} and a class of fractional quantum Hall states (FQHS) with either an even-denominator filling factor or non-abelian statistics \cite{Halperin1983, Rezayi1989, Yang2008}.

In this paper we explore conditions for the development of pairing correlations in quantum insulators at zero temperature, and their observable consequences. Our analysis is focused on the nearly unitary scattering in the $s$-wave channel, for several reasons. First, the theoretical convenience of studying the unitarity limit is well matched with the present-day cold atom experiments which routinely access unitarity in the $s$-wave channel. Second, demonstrating insulators dominated by $s$-wave Cooper pairs reveals that correlated non-superconducting states with pairing fluctuations are not the privilege of unconventional pairing mechanisms such as found in $d$-wave cuprates. An indirect implication is that certain unconventional properties of normal states in high-$T_c$ superconductors could indeed be the consequence of certain universal aspects of
pairing fluctuations, rather than the specific microscopic mechanisms responsible for pairing. In this manner, our results indirectly support the ``charge-oriented'' or ``vortex liquid'' picture of cuprates, with an added benefit of taking into account the fermionic excitations. The $s$-wave Cooper pair insulator may even be susceptible to charge density wave order as we shall briefly discuss, which is another interesting similarity to cuprates.
Finally, by focusing on the $s$-wave channel, we can sidestep the complications
that are associated with gapless fermions in nodal $d$-wave superconductors and
the closely related issue of the absence of true "real space" pairs -- and
thus of the BEC counterpart to the BCS regime -- in
systems where the pairing correlations are intrinsically repulsive. In those
circumstances, strong quantum fluctuations ultimately lead to break-up of Cooper
pairs and the low-energy physics of ensuing "d-wave duality" demands description
which goes beyond the quantum
liquid of vortex-antivortex excitations \cite{Tesanovic2008}. Nevertheless,
even in such systems, an intermediate state -- having strong pairing
correlations but lacking the long-range superconducting order -- is typically
present in the phase diagram,
and many of its properties will reflect the universal aspects of our $s$-wave
insulator \cite{Tesanovic2008, tesanovic04}.

A sharp statement which follows from renormalization group (RG) \cite{Nikolic2010} and arguments in this paper is that precursor $s$-wave pairing in the insulating ground state is tied to the BEC regime which is separated from the weak-coupling BCS regime by a scattering resonance. Similar pairing phenomena have been studied in the deep quantum Hall limit \cite{nikolic:144507} and at finite temperatures \cite{Stajic2004, Chien2008a}. Presently we view the BEC and BCS regimes as effective and appropriate for a particular band insulator or other gapped state of fermionic quasiparticles. Since the lattice potential frustrates the motion of microscopic particles, the strength of microscopic interactions is effectively enhanced to pull the system toward the effective BEC limit. The RG indicates that a BEC-BCS crossover with attractive interactions can be observed only above two dimensions, where the BEC regime requires a sufficiently strong coupling. In two dimensions, however, even arbitrarily weak coupling puts the system into the BEC regime, implying the existence of charged bosonic excitations in the vicinity of the superconducting transition. The second order superconductor-insulator transition in a BEC regime is captured by a purely bosonic effective theory, as the fermionic excitations remain gapped across the transition. Therefore, the insulating state is a bosonic Mott insulator, which can be also viewed as a relativistic quantum liquid of vortices and antivortices if the transition is driven at a fixed quasiparticle density in the grand-canonical ensemble.


A clear and observable manifestation of a bosonic Mott insulator adjacent to the superconducting phase is the bosonic mean-field or XY universality class of the transition. However, a Mott insulator should then have a more robust identity, something to distinguish it from a band insulator which can be turned into a superconductor only by closing the fermion gap. Here we construct a special spectral function which captures non-analytic changes of the many-body spectrum associated with the disappearance of excitations with bosonic statistics. This function behaves as an order parameter at a second order transition when the insulator changes its character from Mott to band. However, no thermodynamic second order phase transition takes place, since only the \emph{excitation} spectrum evolves in a non-trivial way. This kind of sharp changes cannot be observed in equilibrium at zero temperature, but could appear as actual phase transitions if the system is driven out of equilibrium.

A concrete non-equilibrium distinction between bosonic Mott and band insulators that we propose is the ability to function as a Cooper pair laser. A coherent excited state which carries a macroscopic charge or current is possible to generate and sustain in a Mott insulator, but not possible in a band insulator. The spectral function we construct is a representation-independent order parameter of the Cooper pair laser, as this driven condensate spontaneously breaks the ``U(1) symmetry''.

An important part of our analysis is a proof that indeed a two- or three-dimensional insulator in an identifiable BEC regime is characterized by the existence of (i) low energy bosonic excitations and (ii) gapped high energy fermionic excitations. Specifically, the fermion gap remains finite across the transition at which the boson gap vanishes. We show this using a model band insulator of quasiparticles and quasiholes near unitarity as a starting point. We consider both the non-relativistic and relativistic effective bosonic dynamics. The propagator for bosonic excitations is calculated in a perturbative random phase approximation manner from the exact fermion propagators and effective short-range interactions embodied in a renormalized vertex function. The mathematical forms of these propagators and vertices are qualitatively fixed by the fact that the ground state is an insulator. In this sense, we derive exact conclusions about the spectra of bosonic and fermionic excitations, taking into account all fluctuation effects. These findings complement the recent RG analysis \cite{Nikolic2010} which was not able to access states with locally stable Cooper pairs.

While cuprates are fundamentally different than the $s$-wave model we analyze, certain remarkable similarities between them are very much worth exploring. If one focuses on the antinodal regions in the Brillouin zone, the superconducting transition in underdoped cuprates looks just like the superconductor-insulator transition we discuss in the two-dimensional BEC regime. The pairing gap is indeed rooted in the antinodal regions where short-range spin correlations alone would gap-out fermionic excitations. The existence of gapless nodal quasiparticles, which prohibits pair bound states and thus spoils the full BEC analogy, is almost a secondary effect of the $d$-wave pairing. What is not spoiled is the dominance of bosonic fluctuations in the pseudogap state, owing to the linearly vanishing density of states of the nodal quasiparticles.

This paper is organized as follows. In section \ref{secBoseIns} we define the model of a band insulator with attractive interactions and discuss its insulating states. Following the derivation of low energy effective theories for the insulating states of our model, we explain in detail the sharp distinction between Mott and band insulators and ways to observe it out of equilibrium. In section \ref{secNumerics} we present an estimate of the Cooper pair insulator regime in the phase diagram of a realistic two-dimensional cold-atom system using a numerical mean-field calculation. The section \ref{secStability} deals with the stability of Cooper pair insulators and presents the perturbative fluctuation analysis. Conclusions are summarized in the discussion section, including experimental prospects and a more detailed qualitative view of the pseudogap physics in cuprates.

\section{Bose v.s. Fermi insulator}\label{secBoseIns}

The most promising circumstances for the occurrence of strongly correlated states can be found whenever there is a macroscopic degeneracy of states which can be lifted even by weak perturbations. One can ask what happens to an integer quantum Hall insulator when electrons acquire attractive interactions. Sufficiently strong attractive interactions in this topological band insulator can yield Cooper pairing and re-entrant superconductivity at zero temperature. However, direct BCS-like pairing (or pair-breaking) transitions are preempted by first order transitions and masked by intervening strongly correlated insulating phases \cite{nikolic:144507}. Such correlated insulators are obtained from superconductors by quantum vortex lattice melting, and hence can be called ``vortex liquids''. While their properties are generally not universal, their existence at zero temperature provides a sharp thermodynamic distinction between paired and unpaired non-superfluid states when time-reversal symmetry is violated.

Here we explore whether some sharp distinction between paired and unpaired insulators can be found in generic time-reversal invariant circumstances. A band insulator with attractive interactions is indeed susceptible to symmetry breaking involving phase-incoherent Cooper pairs. A recent study has shown that pairing instabilities in ideal multi-band insulators \emph{always} occur at finite rather than zero pair-momenta. Such instabilities produce \emph{supersolid} pair density wave (PDW) states \cite{Nikolic2009a}. The pairing wavevector is generally incommensurate with the lattice as a consequence of the non-trivial crystal-momentum dependence of vertex functions for attractive contact interactions. While fluctuations cannot eliminate the finite-momentum instability, a simple argument shows that they could have a profound influence on selecting the wavevector. The incommensurate supersolid PDW, which naively results from this instability, would be highly frustrated in the lattice potential. The resulting vast multitude of degenerate PDW ground-states can be easily mixed by quantum fluctuations until the U(1) symmetry is restored, or a supersolid at a commensurate wavevector is stabilized. We will show in this paper that strong quantum fluctuations near the second order transition can result with a bosonic Mott insulator of Cooper pairs which intervenes between the superconductor and band insulator states in the phase diagram. Then, the fluctuations which stabilize a commensurate order in the supersolid phase are expected to do the same in the Mott insulator, and produce a commensurate \emph{insulating} PDW. The obtained insulating phase is thermodynamically distinct from any featureless band insulator due to translation symmetry breaking. It is also distinguished from ordinary ``single-fermion'' charge density waves by its low energy charged bosonic excitations. Translational symmetry breaking is possible even though the number of Cooper pairs per site is an integer, due to the orbital degrees of freedom carried by Cooper pairs, and peculiar non-analytic features of bosonic propagators and vertices at the zero momentum \cite{Nikolic2009a}.

Most generally, however, the insulating state proximate to the superconductor need not have broken symmetries, or any other thermodynamic distinction from the ordinary band insulator. Nevertheless, it is still possible to sharply distinguish it from the band insulator in out-of-equilibrium situations. We will argue that our model indeed supports an insulator with infinite-lifetime Cooper pairs. The physical picture is as follows. While the fermionic excitations have a large bandgap, bosonic collective excitations must have a much smaller gap in the insulator near the second-order superconducting transition. The conservation of energy, momentum and charge protects the infinite life-time of these bosons (in an isolated system at $T=0$) despite interactions among them. While some relaxation processes can take place resulting with decays into fermion pairs, excited momentum and charge carrying collective modes below the bandgap cannot completely vanish without coupling to the environment. Therefore, by coupling to a suitable external driving system, it is possible, as a matter of principle, to excite this bosonic insulator into a macroscopically coherent state which spontaneously breaks U(1) symmetry. Such a driven (non-relativistic) condensate is analogous to the state of (relativistic) photons produced by a laser. If the boson gap is pushed above twice the fermion gap, for example by reducing the strength of pairing interactions, then the coherent bosonic excitations cease to exist and the boson-laser non-equilibrium state becomes impossible.

Formally, we define phase transitions as points in the parameter space $\lbrace \mu_i, \nu_i \rbrace$ at which the thermodynamic potential $\mathcal{F}(\mu_i,\nu_i)$ is not an analytic function. If we know the Hamiltonian $H(\mu_i)$, we can construct the density matrix $\rho$ as a function of both microscopic parameters $\mu_i$ and external parameters $\nu_i$ which can describe the environment (temperature, details of the non-equilibrium drive, etc.). In thermal equilibrium at temperature $T=1/\beta$ we simply have $\rho = z^{-1} e^{-\beta H}$, where $z=\Tr{tr}(e^{-\beta H})$, and $\mathcal{F}=-T\log z$. Clearly, at $T=0$ only the ground-state manifold of the Hamiltonian enters $\mathcal{F}$, so that quantum phase transitions are possible only if the ground-state spectrum undergoes non-analytic changes, such as changes of degeneracy. However, the density matrix at finite temperatures, or in generic non-equilibrium situations (even at $T=0$) can inherit non-analytic behavior from the excited state spectrum. Deriving thermodynamic functions in these general cases provides mathematical routes for defining ``non-equilibrium phase transitions'' and sharp distinctions between states which may be thermodynamically equivalent. The Cooper pair insulator described above is an example of this principle.

\subsection{The model and insulating states}

As a generic model of a band insulator let us consider the imaginary-time action of neutral fermionic particles with attractive interactions $U$, in a lattice potential $V(\rv)$:
\begin{eqnarray}\label{ContModel1}
&& S = \int \dd\tau \biggl\lbrack \dd^{d}r \psi_{\alpha}^{\dagger}
  \left( \frac{\partial}{\partial\tau} - \frac{\nabla^2}{2m} + V(\boldsymbol{r}) - \mu \right) \psi_{\alpha} \\
&& -\int \dd^{d}r_1 \dd^{d}r_2 U(|\rv_1-\rv_2|)
       \psi_{\uparrow}^{\dagger}(\rv_1) \psi_{\uparrow}^{\phantom{\dagger}}(\rv_1)
       \psi_{\downarrow}^{\dagger}(\rv_2) \psi_{\downarrow}^{\phantom{\dagger}}(\rv_2) \biggr\rbrack \nonumber
\end{eqnarray}
This is a microscopic multi-band model defined in continuum space and not a priori tied to the vicinity of a critical point. Let us imagine that we can control the chemical potential $\mu$, lattice $V(\rv)$ and the overall strength of attractive interactions $U$ between fermions. In cold-atom realizations of the model it is easy to manipulate $U$ through Feshbach resonances and the depth of an optically generated lattice potential, while condensed matter systems allow direct control of the chemical potential.

Suppose we place the chemical potential somewhere in a bandgap of the fermion spectrum and start with strong attractive forces for which the ground state is a superfluid. Then, we gradually reduce $U$ at zero temperature. By focusing on three dimensions and not too shallow lattice potentials we can ensure the following changes. First, the system undergoes a transition to an insulator at some critical $U=U_{sf}$. Goldstone modes of the superfluid acquire a gap, but initially this gap is much smaller than the gap for fermionic excitations. In this regime, the low-energy effective theory can be obtained from (\ref{ContModel1}) by applying a Hubbard-Stratonovich transformation in the particle-particle channel to decouple the interaction, followed by the fermion field integration. The lattice theory which describes the superconducting transition, derived in Appendix \ref{appBoseModels}, contains only charged bosonic fields (Cooper pairs):
\begin{equation}\label{EffModel2b}
S_{\Tr{eff}}^{(2)} = \int \dd\tau \biggl\lbrace \sum_{ij} b_i^{\dagger}
    K_{ij}^{\Tr{eff}}\left[\frac{\partial}{\partial\tau}\right] b_j^{\phantom{\dagger}}
  + \sum_{ijkl} \mathcal{U}_{ij}^{kl} b_i^{\dagger} b_j^{\dagger}
    b_k^{\phantom{\dagger}} b_l^{\phantom{\dagger}} + \cdots \biggr\rbrace \ .
\end{equation}
The indices $i,j\dots$ label both the lattice sites and orbital degrees of freedom, and $K_{ij}^{\Tr{eff}}[\hat x]$ is an operator function of $\hat x$ with a well defined Taylor expansion about $\hat x=0$. Note that all terms in this action have local character owing to the finite gap of the fermionic fields whose fluctuations produce the effective boson dynamics. This means that the momentum $\pv$ and frequency $\Omega$ dependence of all couplings is analytic at $\pv=0$ and $\Omega=0$, so that the bosonic fields $b$ correspond to physical bosonic excitations even though their mathematical origin is in a Hubbard-Stratonovich transformation. The insulator adjacent to the superconductor is essentially a bosonic Mott insulator.

As $U$ is decreased even further, the mass terms for all boson fields grow. The boson gap increases and the energy range spanned by the long-lived bosonic excitations keeps shrinking, being limited from above by the threshold for decay into fermions. At some other value $U=U_{bi}$ the boson gap becomes equal to twice the fermion gap. Beyond that point, the lowest energy excitations are not bosonic, and the above effective theory is not applicable. Instead, the theory which captures dynamics at the lowest energies can be obtained by writing the multi-band lattice version of the single-channel model (\ref{ContModel1}), or formally by integrating out the $b$ fields in (\ref{LatticeModel1}) or (\ref{LatticeModel2}):
\begin{eqnarray}\label{EffFermion}
S_{\Tr{eff}}^{(\Tr{bi})} & = & \int \dd\tau \biggl\lbrack \sum_i
        f_{i\alpha}^{\dagger} \left( \frac{\partial}{\partial\tau} - \mu \right) f_{i\alpha}^{\phantom{\dagger}}
        -\sum_{ij} t_{ij}^{\phantom{\dagger}} f_{i\alpha}^{\dagger} f_{j\alpha}^{\phantom{\dagger}} \nonumber \\
  & - & \sum_{ijkl} U_{ijkl} f_{i\uparrow}^{\dagger} f_{j\uparrow}^{\phantom{\dagger}}
                             f_{k\downarrow}^{\dagger} f_{l\downarrow}^{\phantom{\dagger}}
  +\cdots \biggr\rbrack \ , ~~
\end{eqnarray}
This theory describes a band insulator.

The theory (\ref{EffModel2b}) has only bosonic excitations at low energies, while (\ref{EffFermion}) has only fermionic excitations. The change of statistics implies that the many-body spectrum cannot smoothly evolve between these two limits. One way to see the non-analytic change is from the spectral function of ``charged'' bosonic excitations, analyzed in section \ref{secStability}, which loses a sharp coherent peak between the two limits. However, we shall pursue a different route in the next section, and consider the density of many-body states in order to discuss implications for out-of-equilibrium dynamics.

\subsection{A sharp distinction between Mott and band insulators at $T=0$}

Let us now start from a bosonic Mott insulator in the limit $0 < \Delta_{\Tr{b}} \ll 2\Delta_{\Tr{f}}$, where $\Delta_{\Tr{b}}$ and $\Delta_{\Tr{f}}$ are the gaps for bosonic and fermionic excitations respectively. The elementary low energy excitations are gapped bosonic modes labeled by crystal wavevector $\kv$ and an orbital index which will be suppressed. The many body spectrum is further characterized by integer numbers of quanta $N_{\kv} \ge 0$ excited in different bosonic modes.

Since all effective interactions between bosons are mediated by gapped fields, they are short-ranged in real space. Consequently, interaction effects become noticeable in the many-body spectrum only when the density of bosonic excitations becomes finite, that is when the average real-space distance between them becomes finite. As long as $\sum_{\kv}N_{\kv}$ is not macroscopically large we can write the following expressions for the total energy $E$ and momentum ${\bf P}$:
\begin{equation}\label{SpectNonint}
E \approx \sum_{\kv} N_{\kv} \EB_{\kv} \qquad , \qquad {\bf P} \approx \sum_{\kv} N_{\kv} \pv_{\kv} \ .
\end{equation}
The individual mode energies are $\EB_{\kv} \ge \Delta_{\Tr{b}}$, while the mode momenta $\pv_{\kv}$ can be arbitrarily small. The total energy is conserved, and the total momentum is conserved modulo first Brillouin zone.

Consider the many-body density of states (DOS) as a function of conserved quantities:
\begin{equation}\label{DOS0}
\rho(E, {\bf P}) = \frac{1}{\mathcal{V}} \sum_n \delta(E-E_n) \delta({\bf P} - {\bf P}_n) \ ,
\end{equation}
where $n$ labels all many-body Hamiltonian eigenstates, and $\mathcal{V}$ is the $d$-dimensional system volume. Let us define a modified ``single-mode'' DOS:
\begin{eqnarray}\label{DOS1}
\rho'(\kv) & = & \lim\limits_{\Delta_{\EB}^{\phantom{d}} \to 0} \lim\limits_{\Delta_p^d \to 0}
  \; \frac{1}{\Delta_{\EB}^{\phantom{d}} \Delta_p^d} \;
  \int\limits_{\Delta_{\EB}^{\phantom{d}}} \dd\delta\EB \int\limits_{\Delta_p^d} \dd^d \delta p \nonumber \\
&& \times \frac{1}{\mathcal{V}} \sum_N \rho(N\EB_{\kv} + \delta\EB, N\pv_{\kv} + {\bf\delta p}) \ .
\end{eqnarray}
The small energy and momentum-space volumes, $\Delta_{\EB}^{\phantom{d}}$ and $\Delta_p^d$ respectively, should be taken to zero only after taking the thermodynamic limit $\mathcal{V}\to\infty$. This ensures that the above energy and momentum integrals of the many-body DOS (\ref{DOS0}) converge in the thermodynamic limit, as the number of modes in a fixed momentum-space volume $\Delta_p^d$ scales as $\mathcal{V}$. The sum over $N$ extracts the essential property of the many-body spectrum, the extent over which the spectrum has the ``period'' $\EB_{\kv}$. The single-mode DOS $\rho'$ would have been divergent even in a finite volume in the absence of interactions between bosons. However, as already pointed out, interaction corrections to spectra scale as $1/\mathcal{V}$ at a fixed density (or $N$), so that \emph{finite} deviations of the many-body spectra from the non-interacting form (\ref{SpectNonint}) at $N\propto\mathcal{V}$ cut off the divergence of $\rho'$. The factor of $1/\mathcal{V}$ in (\ref{DOS1}) then keeps $\rho'$ finite in the thermodynamic limit. Therefore, $\rho'$ will by construction reflect the density of bosonic quanta which can be excited in a single mode before interactions become too strong, which suggests an ``order parameter'' interpretation. We shall immediately see that the behavior of $\rho'$ indeed resembles a second order transition at the crossover between Mott and band insulators.

\begin{figure}
\vskip -0.25in
\subfigure[{}]{\includegraphics[width=1.2in, bb=0 -12 121 80]{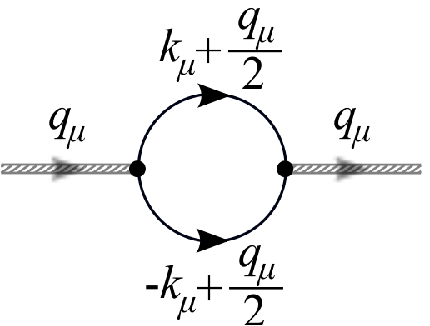}} \hspace{0.1in}
\subfigure[{}]{\includegraphics[width=2.0in, bb=0 30 196 174]{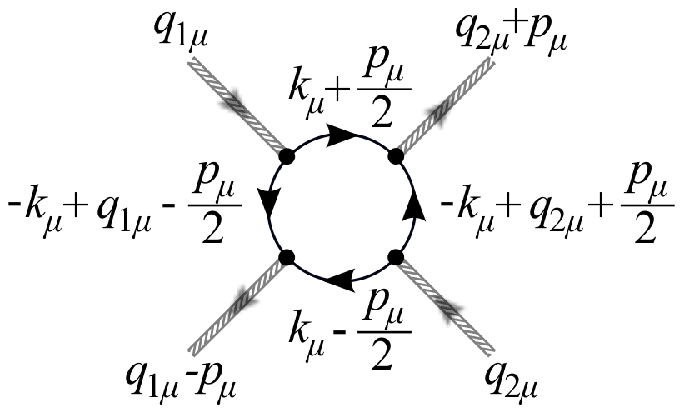}}
\caption{\label{BosonInteraction}Diagrammatic representation of (a) quadratic and (b) quartic terms in the effective action (\ref{EffModel2b}) describing the dynamics of bosonic excitations. Solid lines represent fermion propagators $G(k_{\mu}) = (\omega-\EE_{\kv})^{-1}$, while the faded external lines represent Cooper pair fields. The diagram (a) generates the inverse pairing susceptibility $\Pi(q_{\mu})$ given by (\ref{MatrixK}) and (\ref{ppBubble}), while (b) generates effective interactions between two bosonic excitations. As long as the boson gap $\Delta_\Tr{b}$ is smaller than twice the fermion gap $\Delta_\Tr{f}$ (implicit in $\EE_{\kv}$), the poles of low energy bosons obtained from $\Pi(q_{\mu})=0$ occur at real frequencies, and the interaction (b) between low energy bosonic excitations has a finite range $r \sim \lbrack 4m(2\Delta_\Tr{f}-\Delta_\Tr{b}) \rbrack^{-1/2}$. Beyond $\Delta_\Tr{b} \ge 2\Delta_\Tr{f}$ the interactions would have been effectively long-ranged if the boson poles survived. However, the boson pole is lost in such a BCS regime (see section \ref{secStability}).}
\end{figure}

Tuning the Hamiltonian parameters to increase the bosonic gap results in the gradual reduction of $\rho'$. The excitation density needed to visibly modify the many-body spectrum away from the free-particle form becomes smaller, as the spatial range of effective interactions between low-energy bosons grows. This can be seen from the Feynman diagram in Fig.\ref{BosonInteraction}(b) which captures the two-body interactions between bosonic excitations. Let us focus on low energy bosons in the Mott insulator whose decay into fermions in a collision is prohibited by momentum and energy conservation. Evaluating this diagram (see appendix \ref{appInt}) at zero energy transfer ($p_0=0$) and finite momentum transfer ($p\equiv|\pv|\neq 0$) in the limit of low energy and momenta of incoming particles reveals a scale $r^{-2} = 4m(2\Delta_\Tr{f}-\Delta_\Tr{b})$ which appears alongside $p^2$. The diagram is convergent in the $p\to 0$ limit when $r$ is finite, but becomes infra-red divergent for $p\to 0$ when $\Delta_\Tr{b}=2\Delta_\Tr{f}$. Therefore, $r$ acts as a characteristic length-scale over which the short-range effective interactions between bosonic excitations are extended.

When the boson gap reaches and exceeds twice the gap for fermionic quasiparticles, the sum over $N$ in (\ref{DOS1}) gets cut-off at an $N/\mathcal{V}\to 0$, so that $\rho' \to 0$ in the thermodynamic limit for every single-particle mode. If bosonic excitations could survive, the inter-boson interactions would become effectively long-ranged ($r\to\infty$) and would completely remove the period $\EB_{\kv}$ from the many-body spectrum. However, more importantly, all bosons, including the lowest-energy ones, now inevitably decay into fermion pairs regardless of whether they collide with other bosons or not. We shall find in the section \ref{secStability} that bosonic poles must completely disappear when $\Delta_\Tr{b} \ge 2\Delta_\Tr{f}$, resulting in a BCS regime. The only elementary excitations are fermionic quasiparticles whose spectrum cannot admit multiply occupied single-particle states. The insulator in this regime is a band insulator \cite{footnote, nikolic:134511}.

Therefore, $\rho'(\kv=0)$ is an analytic function of the many-body spectrum in a finite-size system, but undergoes a non-analytic change in the thermodynamic limit at the crossover $\Delta_\Tr{b}=2\Delta_\Tr{f}$ between a bosonic Mott and fermionic band insulator. This non-analytic feature stems from the fundamental change in the many-body spectrum of excited states, and can be observed as a ``second order phase transition'' in out-of-equilibrium circumstances.

A similar non-analytic change occurs in $\rho(\kv)$ for any $\kv$ when $\EB_{\kv}=2\Delta_\Tr{f}$. The abundance of singularities at different $\kv$ does not trivialize the matter since the ensemble averaged operators can pick up different sets of these singularities in different out-of-equilibrium circumstances. For example, $\rho(\kv)$ can be regarded as an order parameter for a driven condensate at wavevector $\kv$, which we discuss in the following subsections. The quantity $\rho(\kv=0)$, however, is special because its non-analytic behavior is the last to disappear when the boson gap is gradually increased, in addition to $\kv=0$ representing the most symmetric excited state. At the equilibrium phase transition to the superfluid ($\Delta_\Tr{b} \to 0$), $\rho(\kv=0)$ becomes proportional to the standard order parameter associated with U(1) ``symmetry breaking''.

\subsection{Cooper pair laser}

A finite $\rho'(\kv)$ in (\ref{DOS1}) is an indication of coherent many-body excitations. Here we argue on more physical grounds that the existence of such states which carry macroscopic charge or current is indeed protected by conservation laws and energetics in the insulating phase with short-range interactions. Therefore, at least in principle, a bosonic Mott insulator can be turned into a Cooper pair laser by some appropriate external drive. The coherent quality of bosonic low energy dynamics is experimentally observable at least out of equilibrium.

Consider preparing an ensemble of $N_\Tr{b}$ coherent Cooper pairs, all occupying the same excitation mode with momentum $\pv$ and energy $\EB_{\pv}$. Let there be no external particle reservoir or heat bath in contact with the system. Due to interactions this is not an eigenstate of the many-body spectrum, so particles will be scattered out of the coherent wave. Some of the Cooper pairs can gain enough energy to break into fermion pairs. After many scattering events, a steady state contains $N'_\Tr{b}$ bosons and $N'_\Tr{f}$ fermions, where a finite fraction of the remaining bosons may still participate in a dressed coherent wave.

Being in a bosonic Mott insulating state with $\Delta_\Tr{b}<2\Delta_\Tr{f}$ ensures the existence of infinite-lifetime single-particle bosonic excitations with energy $\Delta_\Tr{b} \le \EB < 2\Delta_\Tr{f}$. However, two such low-energy bosons can still have enough total energy to leave behind a pair of free fermions after a collision as one of the bosons is pushed to high energy above $2\Delta_\Tr{f}$ and the other drops to a lower energy and takes a recoil momentum. Hence, we must first explore whether at least some bosons from the initial ensemble can survive collisions. The conservation of energy and charge requires:
\begin{eqnarray}\label{ECconserv}
N_\Tr{b} \EB_{\pv} & = & N'_\Tr{b} (\Delta_\Tr{b} + K_\Tr{b}) + N'_\Tr{f} (\Delta_\Tr{f} + K_\Tr{f}) \\
2 N_\Tr{b} & = & 2 N'_\Tr{b} + N'_\Tr{f} \nonumber
\end{eqnarray}
Here $K_\Tr{b} \ge 0$ and $K_\Tr{f} \ge 0$ are the amounts of kinetic energy per surviving low-energy boson and excited high-energy fermion respectively, some of which can be considered heat. It follows that
\begin{equation}\label{CPL}
N'_\Tr{b} = \frac{\EB_{\pv} - 2 (\Delta_\Tr{f} + K_\Tr{f})}{\Delta_\Tr{b} + K_\Tr{b}
  - 2 (\Delta_\Tr{f} + K_\Tr{f})} N_\Tr{b} \ .
\end{equation}
Since $N'_\Tr{b} \le N_\Tr{b}$, we must have $\EB_{\pv} \ge \Delta_\Tr{b} + K_\Tr{b}$. If the energy $\EB_{\pv}$ of the initial Cooper pairs is larger than the threshold $2\Delta_\Tr{f}$ for the decay into two fermions, then $K_\Tr{f}$ can grow until $N'_\Tr{b}$ becomes zero, implying the decay of all Cooper pairs. The increase of entropy likely leads to this outcome. Otherwise, $\EB_{\pv}<2\Delta_\Tr{f}$ makes the numerator in (\ref{CPL}) negative (for any $K_\Tr{f} \ge 0$), so that the denominator must be negative as well in order to keep $N'_\Tr{b}$ positive. Then, a consequence of $K_\Tr{b} \ge 0$ is
\begin{equation}
N'_\Tr{b} \ge \frac{1 - \frac{\EB_{\pv}}{2 (\Delta_\Tr{f} + K_\Tr{f})}}
   {1 - \frac{\Delta_\Tr{b}}{2 (\Delta_\Tr{f} + K_\Tr{f})}} N_\Tr{b} \ .
\end{equation}
In other words, $N'_\Tr{b}$ is a finite fraction of $N_\Tr{b}$, unless $\EB_{\pv}$ reaches or exceeds $2 (\Delta_\Tr{f} + K_\Tr{f})$ when $N'_\Tr{b}$ can drop to zero. Therefore, if a bosonic mode with energy $\Delta_\Tr{b} \le \EB_{\pv} < 2 \Delta_\Tr{f}$ is macroscopically excited, then the resulting steady state will still have a macroscopic number of surviving bosons.

The conservation laws merely protect the ensemble against losing all bosons to fermions. However, we must yet argue that a finite fraction of the remaining bosons participates in a macroscopic coherent wave. This is in fact straight-forward and on the same footing as the case for the stability of equilibrium condensates. Let us boost to the ``center-of-mass'' reference frame of the ensemble where a coherent wave would look like a regular static condensate. If no bosons were to remain in the coherent wave due to collisions, then they would have to be localized into a ``crystal'' in this frame at distances $a\propto n^{-\frac{1}{d}}$ from each other, where $n$ is their density. By Heisenberg uncertainty principle the localization (kinetic) energy per particle would scale as $a^{-2}$, while the potential energy due to short-range interactions can be modeled as an exponential-like function $e^{-\kappa a}$. Clearly, regardless of the numerical co-efficients in these functional forms, localization becomes too costly at large enough $a$, so the ``crystal'' melts at a finite low density $n$. The neglect of the lattice potential for the purposes of this argument is justified in the limit of $n$ much smaller than one particle per site.

Therefore, coherent many-body excitations are stable if they hold a macroscopic but not too large number of Cooper pairs. Relaxing the ensemble to a maximum entropy state with the given macroscopic momentum (by the external Cooper pair laser pump) would leave behind a macroscopic number of bosons in a coherent wave because this minimizes the energy of particles which do not take away heat.


Another possible issue are finite wavevector dynamical instabilities of the excited coherent wave \cite{Wu2001, Altman2005, Polkovnikov2005, Burkov2008, Ganesh2009}. However, we do not expect such instabilities to arise near conventional second order superconductor-insulator transitions that we focus on. Since the equilibrium Mott insulator has an integer number of Cooper pairs per site, its lowest energy excitations live at the ordering wavevector(s) of the adjacent superconducting state (uniform or PDW). Any roton excitations at other crystal wavevectors are separated by an additional gap, so they do not provide a channel for dynamical instabilities at slow excited flows. Note that the Cooper pair laser is fundamentally dynamically unstable at zero \cite{Altman2005, Polkovnikov2005, Mun2007} or any PDW wavevectors - this is precisely what the ``laser pump'' has to defeat in order to drive a coherent condensate.

\subsection{A non-equilibrium $T=0$ phase diagram from a hypothetical experiment}

While many out-of-equilibrium phase transitions have been considered in literature, the overall theory is far less developed than in the case of equilibrium (thermodynamic) phase transitions. Systematic studies of critical dynamics, expressed through properties such as relaxation rates and transport, elucidated a number of non-equilibrium universality classes \cite{Hohenberg1977}. In many cases critical dynamics was analyzed in the vicinity of equilibrium critical points \cite{damle97, sengupta04, Podolsky2007, Herzog2007, Hartnoll2007}. Also, very detailed studies of non-equilibrium physics have been devoted to one-dimensional systems \cite{Barouch1971a}, where an additional incentive comes from integrable cases which lack ergodicity and do not reach thermal equilibrium if decoupled from the environment. This interest in one-dimensional systems has recently grown in the context of ultra-cold atoms \cite{Greiner2002a, Weiss2006, Hofferberth2007, Burkov2007}. On the experimental side, various demonstrations of phase transitions caused by driving a system out of equilibrium have been demonstrated \cite{Ogasawara2000, Mani2002, Iwai2003, Perfetti2006}. A more complete phase diagram is parametrized by both equilibrium and non-equilibrium parameters, so that the same transition can be driven either in equilibrium or out of equilibrium \cite{Takei2008}.

\begin{figure}
\includegraphics[width=2.5in]{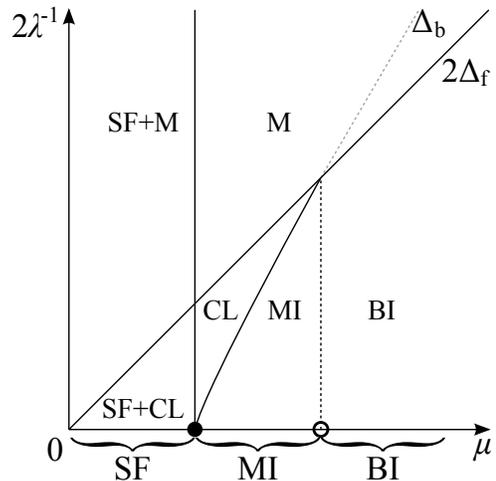}
\caption{\label{NoneqPD}A schematic zero-temperature phase diagram of the model (\ref{ContModel1}) driven out of equilibrium by a laser of wavelength $\lambda$ ($c=\hbar=1$). We assume that the laser induces at least two-photon processes in which Cooper pairs are promoted to higher energy states. In equilibrium at $\lambda^{-1}=0$ there are two thermodynamic phases, superfluid (SF) and insulator separated by a critical point (filled circle). The many-body spectrum undergoes another non-analytic change in the thermodynamic limit (open circle) which affects excitations and introduces a sharp distinction between a bosonic Mott insulator (MI) and a fermionic band insulator (BI). This ``transition'' may not be thermodynamic, there is no latent heat, onset of symmetry breaking or topological order. Going out of equilibrium allows dynamic population of higher energy states at expense of depleting the ground state. If the excited states have bosonic character, light absorption can yield a driven condensate, or a Cooper pair laser (CL). However, if the two-photon energy exceeds twice the fermion gap, the excited Cooper pairs decay into free quasiparticles and a driven metal (M) is obtained. When the ground-state is a superfluid, the driven excited state can coexist with the depleted ground-state superfluid (SF+CL and SF+M). The labels $\Delta_{\Tr{b}}$ and $\Delta_{\Tr{f}}$ denote gaps for bosonic and fermionic excitations, which are controlled by the chemical potential $\mu$. The vertical dashed line indicates the non-analytic change of the spectrum as a function of only the equilibrium parameters, which may not be felt within the ground-state manifold.}
\end{figure}

Here we attempt to put the previous discussion in the context of a fictitious experiment which aims to probe the boson-dominated dynamics in the insulating state. We shall imagine that a laser light of wavelength $\lambda$ which can excite collective bosonic modes above the ground state is shined on the sample. For this purpose we are interested in absorptions of an even number of photons in which there is no angular momentum transfer (the laser should be linearly polarized). As long as the momentum of absorbed photons is negligible, excitations beyond the absorption edge must contain more than one bosonic mode. Single photons can also be absorbed, but the threshold for such processes is given by twice the fermion gap and lies above the threshold for the two-photon absorption in the interesting Bose insulator regime. The schematic phase diagram is shown in Fig.\ref{NoneqPD}.

If the ground-state is a superfluid (SF), any laser wavelength $\lambda$ will result in some partial depletion of the ground-state condensate and promotion of some Cooper pairs to finite energies. If the energy of promoted Cooper pairs is larger than twice the fermion gap $2\Delta_{\Tr{f}}$, the excited pairs decay and leave behind a population of free fermions which can conduct current as a metal. Otherwise, the promoted Cooper pairs can survive as long-lived low energy excitations and form a coherent state, a Cooper pair laser (CL).

If the ground-state is an insulator, it may still be possible to obtain a Cooper pair laser, as long as the excited low-energy Cooper pairs cannot decay into fermions. This possibility exists if the boson gap $\Delta_{\Tr{b}}$ is smaller than twice the fermion gap $2\Delta_{\Tr{f}}$. Otherwise, photon absorption yields an excited population of interacting but free fermions. The insulating ground-state which can be excited into a driven condensate is designated a bosonic Mott insulator (MI) in Fig.\ref{NoneqPD}. The analysis in the previous subsection points to the conclusion that the many-body spectrum of the \emph{equilibrium} Hamiltonian undergoes a qualitatively significant non-analytic change when $\Delta_{\Tr{b}}=2\Delta_{\Tr{f}}$, the open circle in Fig.\ref{NoneqPD}. The spectrum remains gapped and it is not clear whether the non-degenerate ground-state energy experiences any non-analytic dependence on the equilibrium parameters. Therefore, we cannot claim any thermodynamic phase transition between the Mott and band insulators, even though formally there exists a sharp distinction between them, visible in the structure of excited many-body states.

Hence, we argue that the phase diagram contains a special MI-BI crossover, depicted by the dashed vertical line in Fig.\ref{NoneqPD}, which has some sharp signature in the spectrum and could be physically described as pair-breaking. The envisioned hypothetical experiment could only indirectly sense the existence of this crossover, by probing the possibility of creating a Cooper pair laser from the insulating equilibrium state.

Realizing Cooper pair laser states in realistic systems is of course very challenging. While dissipation due to absorption can always be defeated by an appropriate laser pump which takes advantage of the stimulated emission, it remains unclear what could be an appropriate pump in a given system, and whether it is feasible in practice (probably not in solids). In trapped cold atom gases, it is at least possible to make an ordinary "condensate" of bosons in an optical lattice, then suddenly change the lattice depth to create conditions for an equilibrium Mott insulator at the trap center. This system, initially in a "laser" state, would take a long time to relax into the many-body ground state. We just do not know how to continuously pump this atomic Cooper pair laser.

\section{Numerical results for a model in the unitarity regime}\label{secNumerics}

Here we estimate the conditions for a bosonic insulator in a specific model, using a mean-field approximation. We consider a model of neutral fermionic ultra-cold atoms in a lattice potential, whose interactions are tuned close to the Feshbach resonance. The imaginary-time ``two-channel'' action describing this system is:
\begin{eqnarray}\label{TwoChannel}
S & = & \int \dd\tau \dd^d r \Biggl\lbrack \psi_{\alpha}^{\dagger}
  \left( \frac{\partial}{\partial\tau} - \frac{\nabla^2}{2m} + V(\rv) - \mu\right)
    \psi_{\alpha}^{\phantom{\dagger}} \nonumber \\
&& + \frac{m\nu}{4\pi}|\Phi|^2 +
     \Phi \psi_{\uparrow}^{\dagger}\psi_{\downarrow}^{\dagger} + h.c. \Biggr\rbrack \ .
\end{eqnarray}
Only contact interactions are relevant near the unitarity fixed point, and they are decoupled by the complex Hubbard-Stratonovich field $\Phi$. We shall use the simplest square lattice potential
\begin{equation}\label{Lattice}
V(\rv) = 2V \sum_{i=1}^d \cos\left( \frac{2\pi x_i}{a_L} \right) \ ,
\end{equation}
where $x_i\in\lbrace x,y,z\dots \rbrace$, $a_L$ is lattice spacing and $V$ is a tunable lattice amplitude. Integrating out the fermion fields $\psi_\alpha$ yields an effective action for the charged bosonic excitations represented by $\Phi$.

\begin{figure*}[!]
\subfigure[{}]{\includegraphics[width=2.2in]{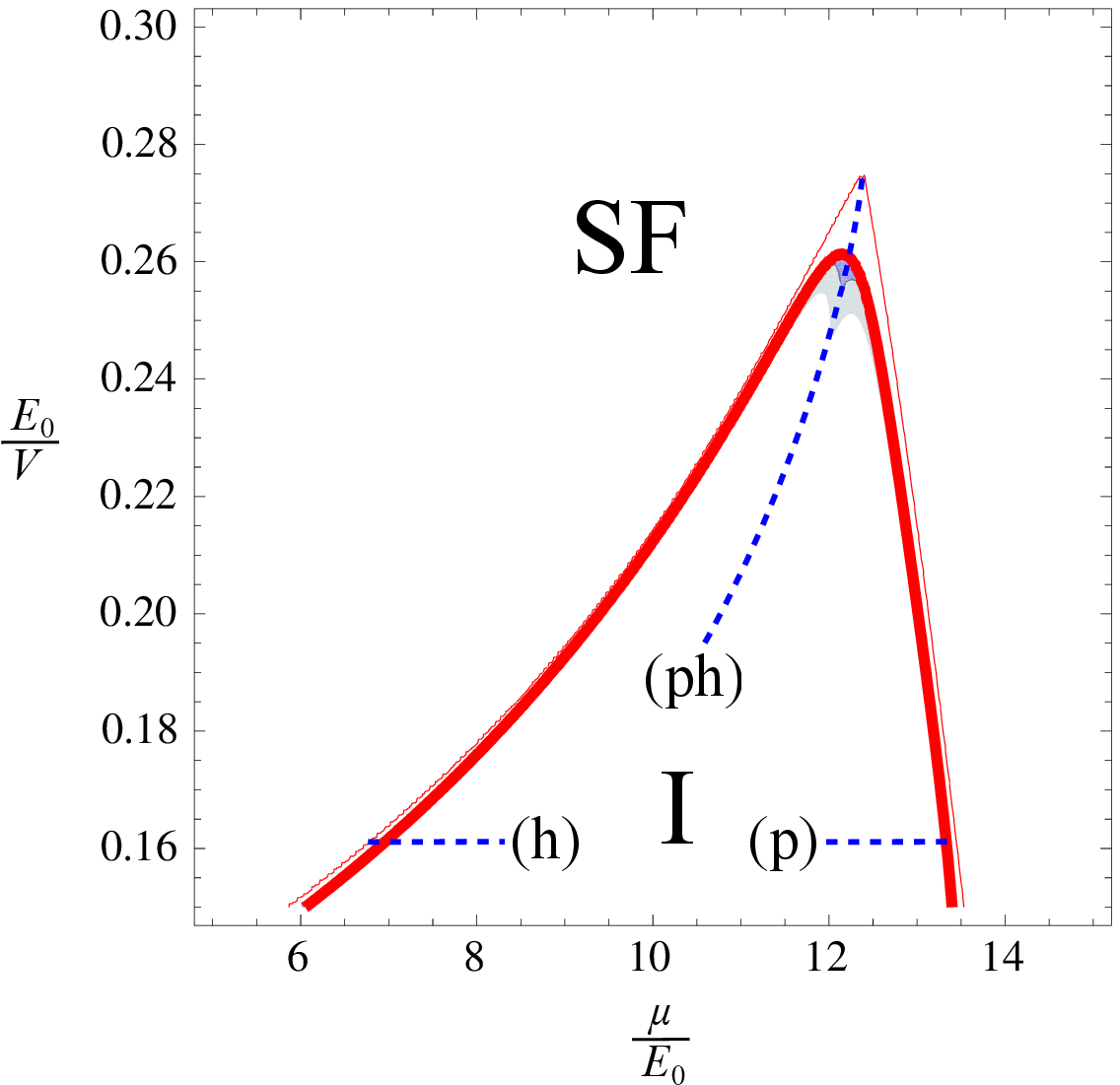}}
\hspace{0.1in}
\subfigure[{}]{\includegraphics[width=2.2in]{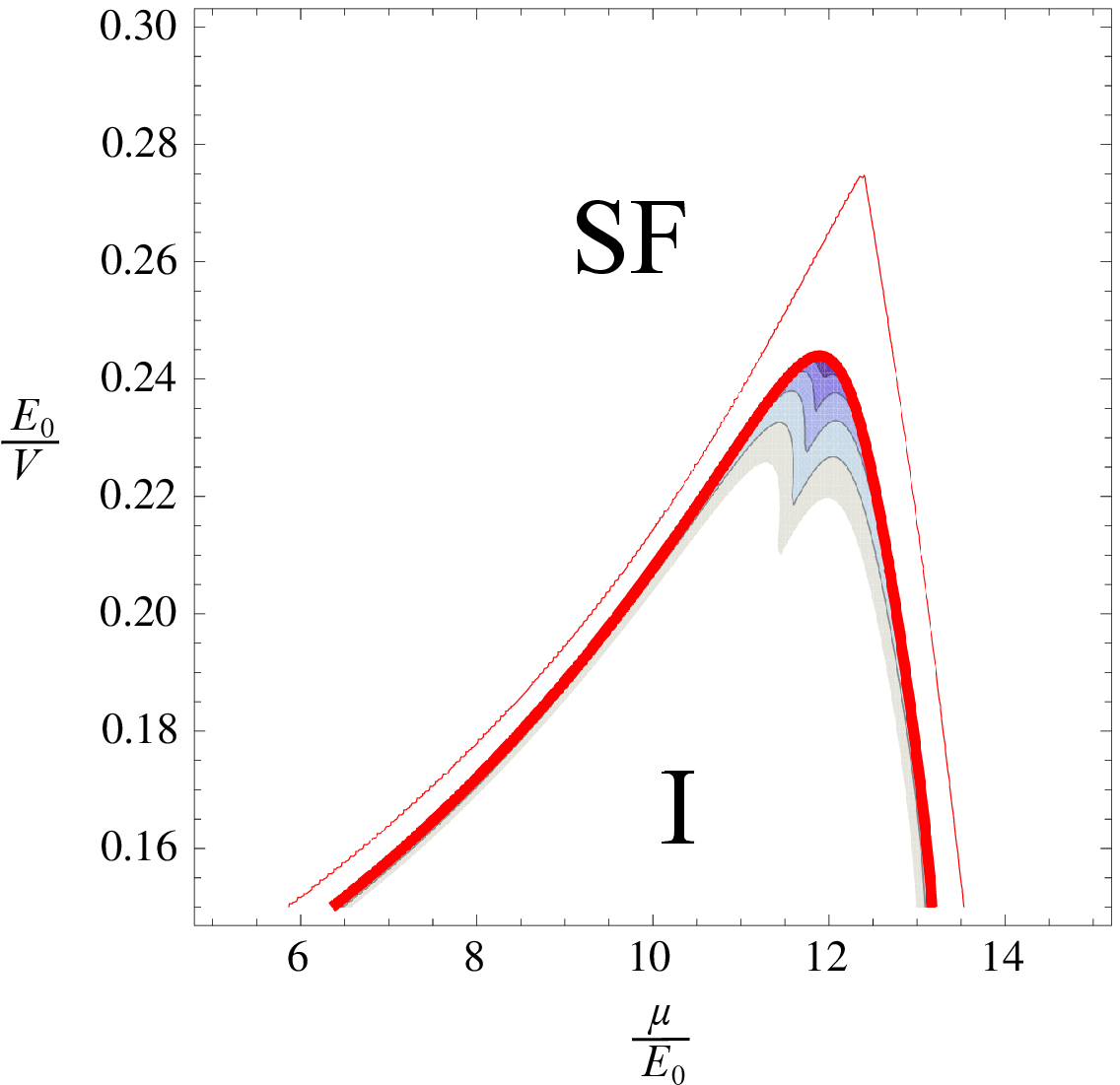}}
\hspace{0.1in}
\subfigure[{}]{\includegraphics[width=2.2in]{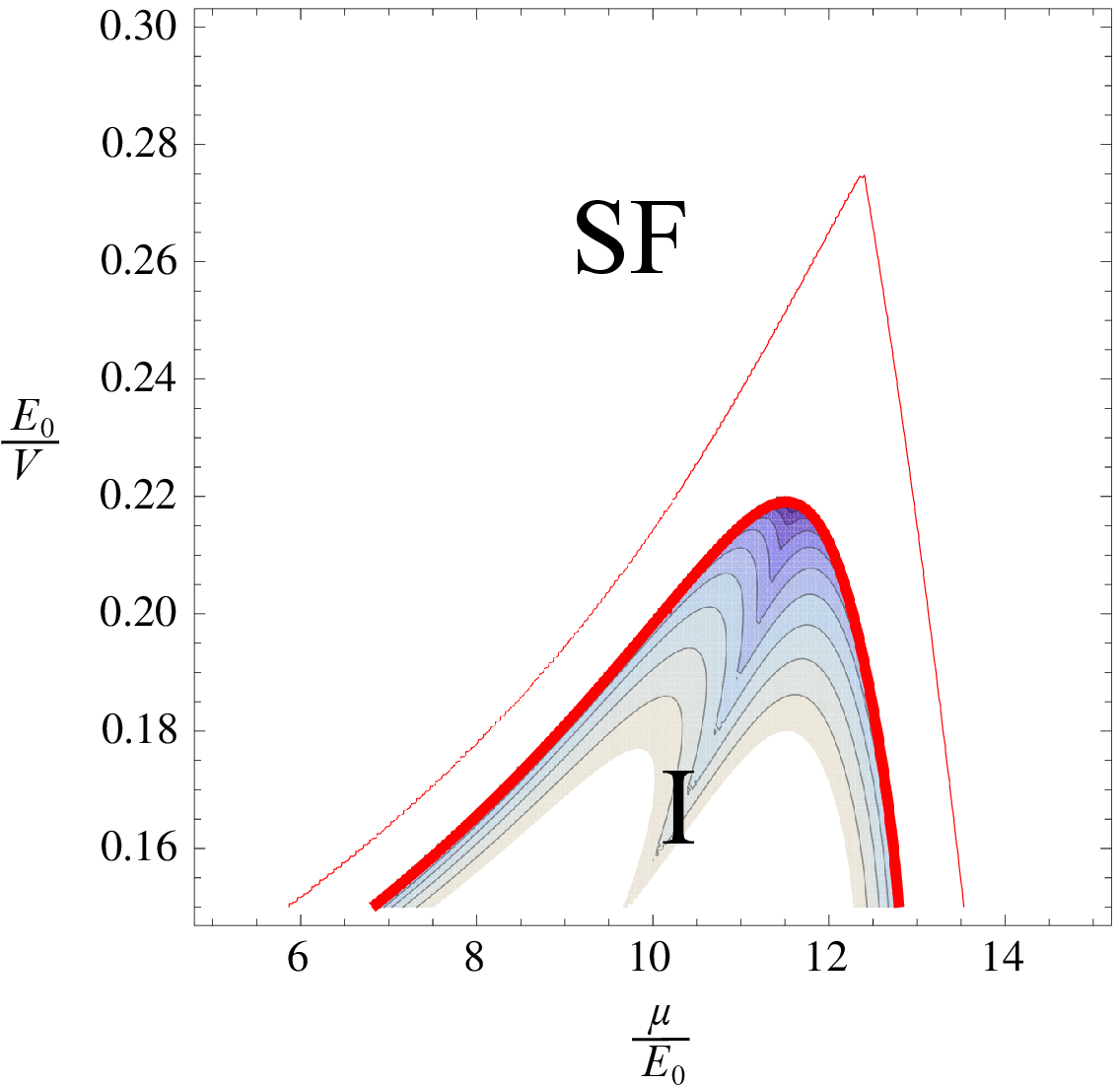}}
\caption{\label{pd1}(color online) The phase diagram of neutral fermions with near unitary interactions, in a two dimensional lattice potential at $T=0$. Going from weak to strong coupling: (a) $\nu=0.5$, (b) $\nu=0$, (c) $\nu=-0.5$ ($E_0=\hbar^2/2ma_L^2$). The red thick line shows the boundary between the superfluid (SF) and insulating (I) phase. The thin red line outlines the first bandgap of bare fermions, toward which the SF-I transition line converges in the vanishing coupling limit $\nu\to+\infty$. The shaded blue regions on the insulating side near the transition represent the difference between the boson and twice the fermion gap, darker colors corresponding to larger differences. These regions which grow with the strength of pairing are Cooper pair insulators. The dashed blue lines in subfigure (a) are trajectories in the parameter space along which transitions dominated by particles (p), holes (h), or both (ph) can occur.}
\end{figure*}

The bubble diagram in the particle-particle channel from Fig.\ref{BosonInteraction}(a)
\begin{eqnarray}\label{MatrixK}
&& \Pi_{\nv\qv;\nv'\qv'}(\Omega) = \!\!\!
    \sum_{\mv_1\mv_2} \int \frac{\dd^d k_1}{(2\pi)^2} \frac{\dd^d k_2}{(2\pi)^2} \times \\
&& \qquad \frac{f\left(\EE_{\mv_1 \kv_1}\right) - f\left(-\EE_{\mv_2 \kv_2}\right)}
         {-\Omega + \EE_{\mv_1 \kv_1} + \EE_{\mv_2 \kv_2}} \;
      \Gamma_{\mv_1 \kv_1 ; \mv_2 \kv_2}^{\nv \qv *} \Gamma_{\mv_1 \kv_1 ; \mv_2 \kv_2}^{\nv' \qv'} \nonumber
\end{eqnarray}
is the first step toward finding a superfluid instability. Here $\kv$ and $\qv$ are conserved first Brillouin zone (BZ) wavevectors, $\nv$ and $\mv$ are ``band-index'' quantum numbers for bosonic and fermionic excitations respectively, $\EE_{\mv \kv}$ are bare fermion energies appropriate for the Bloch states (relative to the chemical potential), $f(\EE)$ is the Fermi-Dirac distribution function, and $\Gamma$ are vertex functions which can be derived from the fermionic $\psi_{\mv \kv}$ and bosonic $\Phi_{\nv \qv}$ Bloch-state wavefunctions:
\begin{equation}\label{Ver}
\Gamma_{\mv_1 \kv_1 ; \mv_2 \kv_2}^{\nv \qv} = \int\dd^3 r \; \Phi_{\nv \qv}^*(\rv)
  \psi_{\mv_1 \kv_1}^{\phantom{*}}(\rv) \psi_{\mv_2 \kv_2}^{\phantom{*}}(\rv) \ .
\end{equation}
Since crystal momentum is conserved, $\Pi_{\nv\qv;\nv'\qv'}(\Omega) = \Pi_{\nv\nv'}(\qv,\Omega)\times (2\pi)^d \delta(\qv-\qv')$. The expression (\ref{MatrixK}) is ultra-violet divergent and we must regularize it by removing the contribution of vacuum zero-point fluctuations occurring at arbitrarily short length-scales. After regularization we may write the physical propagator $D_{\nv\nv'}(\qv,\Omega)$ for the charged bosonic excitations:
\begin{equation}\label{BosonProp}
-D_{\nv\nv'}^{-1}(\qv,\Omega) = \frac{m\nu}{4\pi} + \Pi_{\nv\nv'}(\qv,\Omega) - \Pi_\Tr{reg} \ ,
\end{equation}
where $\Pi_\Tr{reg}$ is the smallest eigenvalue of $\Pi_{\nv\nv'}(\qv,\Omega)$ evaluated in vacuum at $T=0$ and $\qv=0$, $\Omega=0$. The formal calculation of $\Pi_{\nv\nv'}(\qv,\Omega)$ in this expression must involve an implicit momentum cut-off $\Lambda$ for internal momenta $\kv_i$, but after regularization (\ref{BosonProp}) is convergent in the $\Lambda\to\infty$ limit. The parameter $\nu$ can be related to the scattering length $a$ in two-fermion collisions in vacuum; in three dimensions, $\nu=-1/a$.

The smallest eigenvalue $\pi(\Omega,\qv)$ of $-D^{-1}_{\nv\nv'}(\Omega,\qv)$ reveals superfluid instability when it becomes negative at zero frequency. Since $\pi(\Omega,\qv)$ is independent of the representation for $D$, we can choose any complete set of boson wavefunctions $\Phi_{\nv\qv}(\rv)$ which treats $\qv$ as a crystal wavevector in the lattice (\ref{Lattice}). In other words, $\Phi_{\nv\qv}(\rv)$ can be Bloch states of any periodic potential with the same symmetries and lattice spacing as (\ref{Lattice}), and it will be convenient to carry out calculations using the plane-wave Bloch states, as if $V=0$, where $\nv\equiv\Gv$ (a reciprocal lattice vector) and $\Phi_{\Gv\qv}(\rv) = e^{i(\Gv+\qv)\rv}$.

In the insulating phase $\pi(0,\qv)>0$, but the boson gap may be small. We estimate the boson gap $\Delta_{\Tr{b}}$ from the fact that $\pi(\Omega,\qv) \approx \pi_{\qv}^{(0)} + \pi_{\qv}^{(1)}\Omega + \pi_{\qv}^{(2)}\Omega^2$ for small boson energy $\Omega$, by noting that $\pi(\Omega,\qv)=0$ yields the dispersion relation for bosonic excitations. In fact, there are two bosonic modes, particle-like with gap $\Delta_\Tr{bc}$ and hole-like with gap $\Delta_\Tr{bv}$:
\begin{eqnarray}
\Delta_{\Tr{bc}} & = & -\frac{\pi_{\qv}^{(1)}}{2\pi_{\qv}^{(0)}} +
  \left\lbrack \left(\frac{\pi_{\qv}^{(1)}}{2\pi_{\qv}^{(0)}}\right)^2 -\frac{\pi_{\qv}^{(2)}}{\pi_{\qv}^{(0)}}
  \right\rbrack^{\frac{1}{2}} \\
\Delta_{\Tr{bv}} & = & +\frac{\pi_{\qv}^{(1)}}{2\pi_{\qv}^{(0)}} +
  \left\lbrack \left(\frac{\pi_{\qv}^{(1)}}{2\pi_{\qv}^{(0)}}\right)^2 -\frac{\pi_{\qv}^{(2)}}{\pi_{\qv}^{(0)}}
  \right\rbrack^{\frac{1}{2}} \nonumber
\end{eqnarray}
The insulator is bosonic if $\Delta_{\Tr{bc}}<2\Delta_{\Tr{c}}$ or $\Delta_{\Tr{bv}}<2\Delta_{\Tr{v}}$, where $\Delta_{\Tr{c}}$ and $\Delta_{\Tr{v}}$ are fermionic particle and hole gaps which we read out from the band-structure at any given chemical potential in this mean-field approximation. This estimate breaks down in the superfluid state.

Figure~\ref{pd1} shows the phase diagram of a two-dimensional system in the vicinity of the superfluid-insulator transition with the average density of two atoms per lattice site. The mean-field approximation finds a band-insulating state sitting inside the first bandgap when the pairing interaction is not strong enough. However, close to the superfluid boundary there is a region in which the gap for collective bosonic excitations is smaller than twice the fermion gap. The thickness of this Cooper pair insulator is small in the weak coupling limit, but grows with the strength of pairing. The spike feature in the shape of the Cooper pair insulator formally reflects the non-analytic dependence of the fermion gap on chemical potential (fermion gap is the distance between the chemical potential and the nearest band-edge for a given lattice depth $V$). Physically, both particles and holes shape the bosonic dynamics in the vicinity of spikes, while only particles or only holes are relevant near the band-edges. We cannot guarantee that such pronounced features would persist beyond the mean-field approximation.

\section{The stability of the Bose-insulator}\label{secStability}

In this section we explore the conditions for having gapped fermionic excitations across the $T=0$ superfluid transition. We are particularly interested in the possibility of ``unconventional'' superfluid transitions across which fermionic excitations remain gapped. The conventional BCS pair-breaking transition has gapless fermions in the normal state, at least just after the transition.

We adopt a simple model of a band insulator in which there are two very broad bands, conduction and valence, separated by a direct bandgap $E_g$. Generalizations to other bandwidths and types of bandgap are straight-forward and are not expected to alter the main conclusions. The chemical potential residing in the bandgap controls the particle and hole gaps $\Delta_\Tr{c}$ and $\Delta_\Tr{v}$ respectively, under the constraint $\Delta_\Tr{c}+\Delta_\Tr{v}=E_g$. The particle energies in the conduction (c) and valence (v) bands are given by
\begin{equation}\label{FermionDisp}
E_\Tr{c}(\kv) = \Delta_\Tr{c} + \frac{k^2}{2m_\Tr{c}} \quad , \quad
E_\Tr{v}(\kv) = -\Delta_\Tr{v} - \frac{k^2}{2m_\Tr{v}} \ .
\end{equation}

It is convenient to carry out calculations in the two-channel model (\ref{TwoChannel}). Restricting ourselves to zero temperature, the task of calculating the bosonic propagator $D$ from (\ref{BosonProp}) in the present model is not complicated. The numerator in (\ref{MatrixK}) is non-zero at $T=0$ only if the two fermions in the bubble diagram are both above or both below the chemical potential. This presently means that only $\mv_1=\mv_2\equiv\Tr{c}$ and $\mv_1=\mv_2\equiv\Tr{v}$ contribute. Assuming a very short-range microscopic interaction potential between fermions, we can ignore the momentum dependence of the vertex functions and write them as $\Gamma_{\mv_1\mv_2}^{\nv}$. Finally, we can choose the representation for $\Pi_{\nv\nv'}$ which diagonalizes it and focus only on the smallest eigenvalue which captures the onset of superfluidity. There is only one set of $\nv$ corresponding to the softest bosonic mode which we ought to focus on, leaving us with only two relevant dimensionless values that the vertex function can take, $\gamma_\Tr{c}=\Gamma_{\Tr{cc}}$ and $\gamma_\Tr{v}=\Gamma_{\Tr{vv}}$.

The Green's function $D(q_\mu)$ for the softest bosonic mode can now be obtained from (\ref{BosonProp}). It will be convenient to express it as $D(q_\mu) \equiv -\pi^{-1}(q_\mu)$, where
\begin{eqnarray}\label{ppBubble}
&& \!\!\!\!\!\!\!\!\!\!\!\!\!\!\! \pi(q_\mu) = \frac{m\nu}{4\pi} +
  \int\frac{\dd^dk}{(2\pi)^d} \;
     \frac{\gamma_\Tr{c}^2}{\Omega-2\Delta_\Tr{c}-\frac{k^2}{m_\Tr{c}}-\frac{q^2}{4m_\Tr{c}}+i0^+} \nonumber \\
&& \!\!\!\!\!\!\!\!\!\!\!\! + \int\frac{\dd^dk}{(2\pi)^d} \;
     \frac{\gamma_\Tr{v}^2}{-\Omega- 2\Delta_\Tr{v}-\frac{k^2}{m_\Tr{v}}-\frac{q^2}{4m_\Tr{v}}+i0^+} - \Pi_\Tr{reg}
\end{eqnarray}
We have parametrized the two-fermion scattering length in the ground state by $\nu$ and a mass scale $m$. Since the ground state is not vacuum but a completely filled valence band, resonant scattering may occur at $\nu \neq 0$, which we shall determine later. We shall also exploit the freedom to define the mass scale $m$ in the most convenient way.

The expression (\ref{ppBubble}) is certainly valid in the saddle-point approximation employing the bare fermion propagators and bare vertices. However, its form does not qualitatively change as a result of quantum fluctuations in the insulating state. One can apply diagrammatic perturbation theory to construct the exact bubble diagram and derive the exact boson Green's function. All excitations, bosonic and fermionic, are gapped, so the propagators and vertices acquire only numerical renormalization. The exact fermion propagators contain dispersions (\ref{FermionDisp}) where the gaps $\Delta_\Tr{c}$, $\Delta_\Tr{v}$ and masses $m_\Tr{c}$, $m_\Tr{v}$ have renormalized values. While these quasiparticles and quasiholes are mixtures of the microscopic particles and holes, their pairing rules are the same as in the saddle-point approximation due to the Fermi-Dirac statistics. Similarly, since the renormalized interactions remain short-ranged, the only relevant consequence of vertex corrections is the renormalization of $\gamma_\Tr{c}$ and $\gamma_\Tr{v}$. Therefore, the exact boson propagator is the same function of renormalized fermion parameters as (\ref{ppBubble}), and we can use it to establish certain relationships between exact bosonic and fermionic excitations in the insulating state.

\subsection{$d=2$}

In two dimensions, integrating momenta up to a cut-off $|\kv|<\Lambda$ in (\ref{ppBubble}) gives the unregularized expression:
\begin{eqnarray}
&& \pi'_{2D}(q_\mu) = -\frac{m_\Tr{c}\gamma_\Tr{c}^2}{4\pi}
  \log\left(1-\frac{\Lambda^2/m_\Tr{c}}{\Omega-2\Delta_\Tr{c}-\frac{q^2}{4m_\Tr{c}}+i0^+}\right) \nonumber \\
&& ~~ -\frac{m_\Tr{v}\gamma_\Tr{v}^2}{4\pi}
  \log\left(1-\frac{\Lambda^2/m_\Tr{v}}{-\Omega-2\Delta_\Tr{v}-\frac{q^2}{4m_\Tr{v}}+i0^+}\right) \ .
\end{eqnarray}
In order to avoid an infra-red divergence, we must assume that fermionic excitations are gapped in the ground state when evaluating $\Pi_\Tr{reg}$. Taking the particle and hole gaps to be $\delta_\Tr{c}$ and $\delta_\Tr{v}$ in the ground state at $\nu=0$ respectively, we find
\begin{eqnarray}\label{ppBubble2D}
&&  \pi(q_\mu) = \frac{m\nu}{4\pi} + \frac{m_\Tr{c}\gamma_\Tr{c}^2}{4\pi}
  \log\left(\frac{2\Delta_\Tr{c}+\frac{q^2}{4m_\Tr{c}}-\Omega-i0^+}{\delta_\Tr{c}}\right) \nonumber \\
&& ~~ +\frac{m_\Tr{v}\gamma_\Tr{v}^2}{4\pi}
  \log\left(\frac{2\Delta_\Tr{v}+\frac{q^2}{4m_\Tr{v}}+\Omega-i0^+}{\delta_\Tr{v}}\right) \ .
\end{eqnarray}
in the $\Lambda\to\infty$ limit.

The boson propagator $D(q_\mu)=-\pi(q_\mu)^{-1}$ has branch-cut singularities at real frequencies $\Omega>\omega_\Tr{c}$ and $\Omega<-\omega_\Tr{v}$ beyond the threshold for Cooper pair decay, where
\begin{equation}\label{omegaPrm}
\omega_\Tr{c} = 2\Delta_\Tr{c}+\frac{q^2}{4m_\Tr{c}} \quad , \quad
\omega_\Tr{v} = 2\Delta_\Tr{v}+\frac{q^2}{4m_\Tr{v}} \ .
\end{equation}
The poles of $D(q_\mu)$ sitting between these branch-cuts on the real axis yield the dispersion of charged bosonic excitations in the insulating state. In order to obtain them from the non-linear equation $\pi(q_\mu)=0$, let us
write
\begin{equation}\label{massPrm}
m_\Tr{c}^{\phantom{2}}\gamma_\Tr{c}^2 = m+\delta m \quad , \quad
m_\Tr{v}^{\phantom{2}}\gamma_\Tr{v}^2 = m-\delta m \ ,
\end{equation}
which defines the masses $m$ and $\delta m$, and separate the terms proportional to $m$ and $\delta m$ in (\ref{ppBubble2D}). We find that $\pi(q_\mu)=0$ is equivalent to
\begin{equation}\label{BetaPrm}
\left(\omega_\Tr{c}-\Omega\right)\left(\omega_\Tr{v}+\Omega\right)
= \delta_\Tr{c}\delta_\Tr{v} e^{-\nu} \left(\frac{\delta_\Tr{v}}{\delta_\Tr{c}} \;
      \frac{\omega_\Tr{c}-\Omega}{\omega_\Tr{v}+\Omega} \right)^{-\frac{\delta m}{m}}
    \equiv \beta^2 \ .
\end{equation}
Therefore, the poles are obtained by solving:
\begin{equation}\label{Poles2D}
\Omega_{1/2} = \frac{\omega_\Tr{c}-\omega_\Tr{v}}{2} \pm \sqrt{\left(\frac{\omega_\Tr{c}+\omega_\Tr{v}}{2}\right)^2
  - \beta^2} \ .
\end{equation}

\begin{figure}
\subfigure[{}]{\includegraphics[height=2.1in]{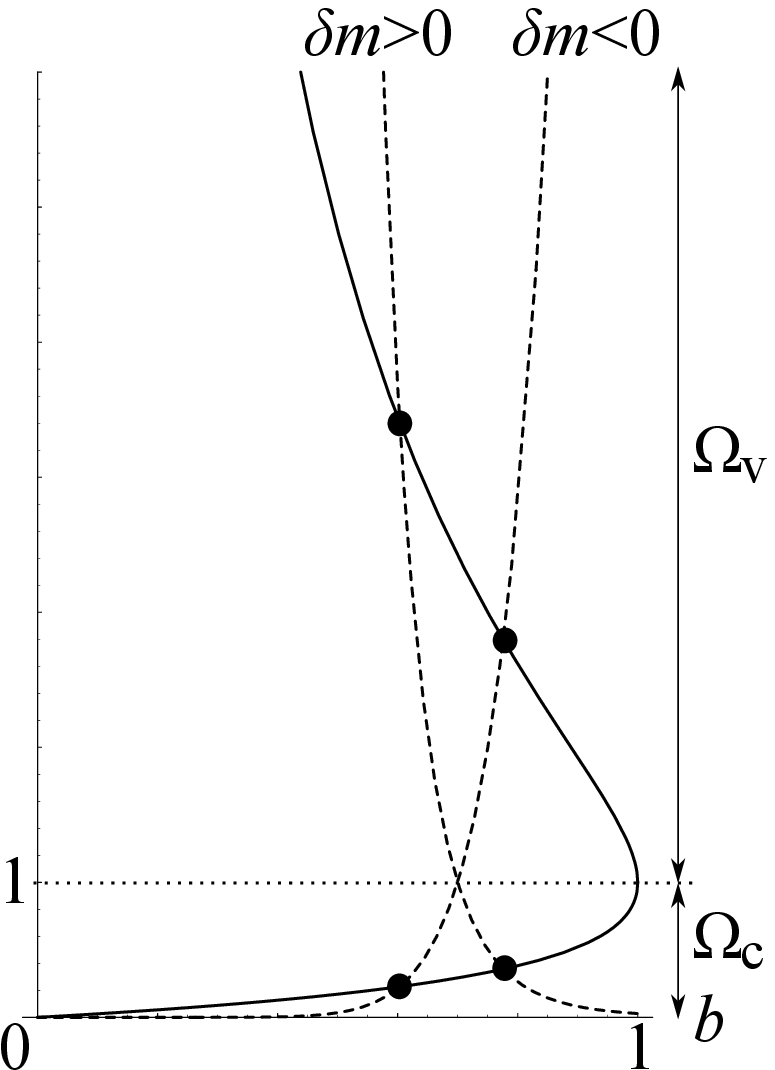}}
\subfigure[{}]{\includegraphics[height=2.0in]{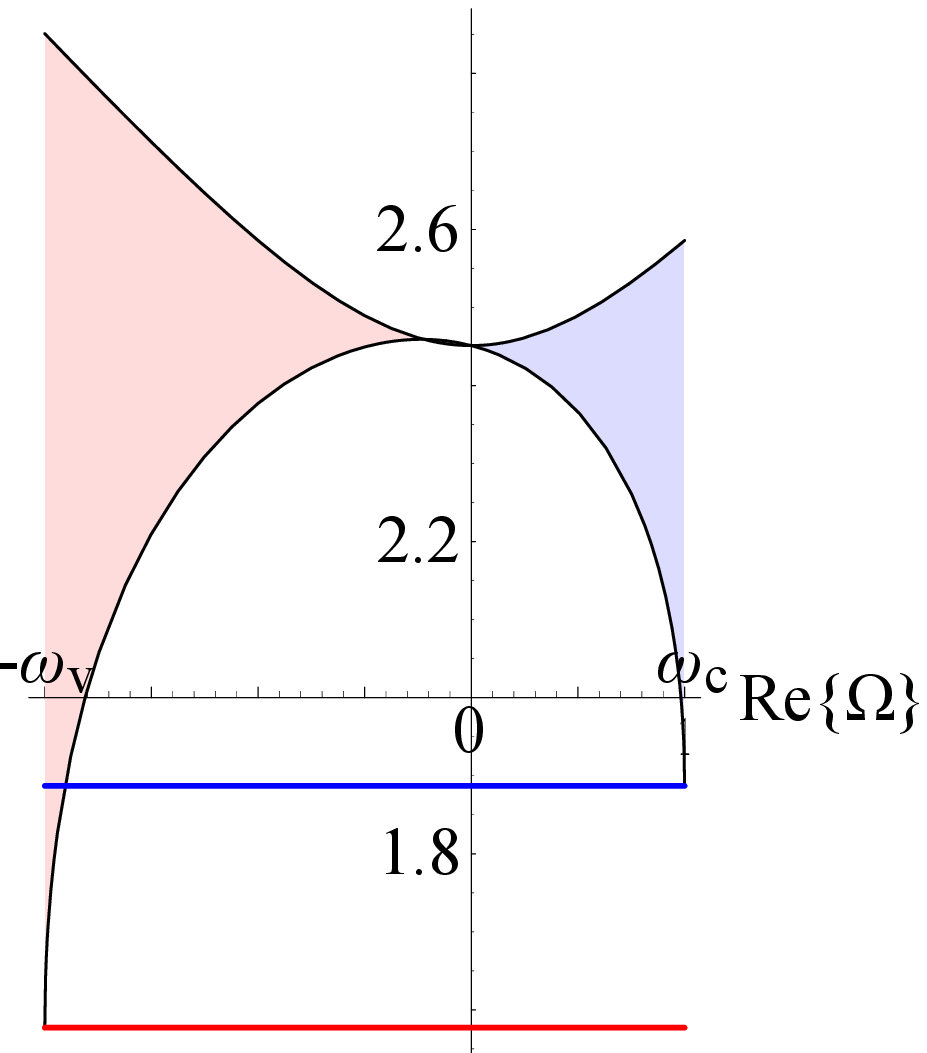}}
\caption{\label{PiSol}(color online) The existence of bosonic poles in two and three dimensions. (a) $d=2$: A plot of $(1\mp\sqrt{1-b})/(1\pm\sqrt{1-b})$ (solid) and $(\delta_\Tr{c}/\delta_\Tr{v})\times(b/b_0)^{-m/\delta m}$ (dashed) from (\ref{2Dsol}) for positive and negative $\delta m$. $b_0=0.7$, $\delta m/m=\pm 0.1$, $\delta_\Tr{c}=\delta_\Tr{v}$ in this example. (b) $d=3$: A plot of $\Tr{Re}\lbrace\pi(q_\mu)\rbrace-m\nu/4\pi$ from (\ref{ppBubble3Db}) as a function of $\Omega\in\mathbb{R}$. The shaded areas are spanned by $\Omega=\Tr{Re}\lbrace\Omega\rbrace e^{i\theta}$ for $\theta\in(-\pi/2,\pi/2)$, the bottom boundary corresponds to $\theta=0$. The two horizontal lines indicate $-m\nu_\Tr{c}/4\pi$ (upper blue) and $-m\nu_\Tr{v}/4\pi$ (lower red). $\omega_\Tr{c}/\omega_\Tr{v}=0.5$, $(\delta m/m)^{3/2}=-0.089$, vertical axis is in arbitrary units in this example.}
\end{figure}

First, we shall verify that the poles always exist by substituting (\ref{Poles2D}) into (\ref{ppBubble2D}). After introducing
\begin{equation}
b = \left( \frac{2\beta}{\omega_\Tr{c}+\omega_\Tr{v}} \right)^2 \quad , \quad
b_0 = \frac{4\delta_\Tr{c}\delta_\Tr{v}e^{-\nu}}{(\omega_\Tr{c}+\omega_\Tr{v})^2} \ ,
\end{equation}
and some algebraic manipulation we obtain the condition equivalent to $\pi(q_\mu)=0$:
\begin{equation}\label{2Dsol}
\frac{\delta_\Tr{c}}{\delta_\Tr{v}} \left(\frac{b}{b_0}\right)^{-\frac{m}{\delta m}} =
  \frac{1\mp\sqrt{1-b}}{1\pm\sqrt{1-b}} \ .
\end{equation}
The Fig.\ref{PiSol}(a) shows plots of the left-hand and right-hand sides of this expression. In general, there are two intersections between these curves for any $b_0$, which can be regarded as a tunable parameter. One intersection corresponds to the particle-like pole $\Omega_\Tr{c}$, and the other one to the hole-like pole $\Omega_\Tr{v}$. In the limit $b_0 \to 0$ the intersections occur at small $b$, so that the right-hand side of (\ref{2Dsol}) behaves either as $b$ or $1/b$, depending on the sign choice, that is the particle/hole nature of the solution. Then, $|\delta m|<m$ guaranties that the curve on the left-hand side always intersects the functions $b$ and $1/b$ on the right-hand side when $b_0 \to 0$, regardless of the sign of $\delta m$. Increasing $b_0$ shifts the intersection points to larger values of $b$, and there is a ``critical'' value of $b_0$ at which the two intersections coalesce and beyond which the poles disappear. However, the superfluid transition occurs before this ``critical'' $b_0$ is reached. Namely, equating the pole frequency to zero at $q=0$ in (\ref{Poles2D}) yields the condition for the superfluid transition:
\begin{equation}\label{PhTrans2D}
4 \Delta_\Tr{c} \Delta_\Tr{v} = \beta^2 \ ,
\end{equation}
which is equivalent to $b = 4 \Delta_\Tr{c} \Delta_\Tr{v} / (\Delta_\Tr{c} + \Delta_\Tr{v})^2 \le 1$. Therefore, the transition occurs before the value of $b$ corresponding to either the particle-like or hole-like pole becomes $1$, which by continuity is before $b_0$ reaches the ``critical'' value. The limiting-case transition with $b=1$ is special and involves condensation into relativistic bosonic modes.

Knowing that bosonic charged excitations always have well-defined particle and hole-like poles in the insulating state, we can examine some of their important properties. We can immediately see from (\ref{Poles2D}) that the bosonic spectrum is relativistic if $\Omega_1 = -\Omega_2$ at $q \to 0$, that is $\Delta_\Tr{c} = \Delta_\Tr{v}$. Equal participation of both particles and holes is required for relativistic dynamics, and the resulting superfluid transition in the XY universality class occurs at the expected place. The particle-hole symmetry is generally lost at finite and especially large $q$, unless $m_\Tr{c}=m_\Tr{v}$. This is natural since the microscopic dynamics is not particle-hole symmetric, but this does not affect the XY universality of the transition, shaped by the long-wavelength modes.

Finally, we observe that the boson gaps are always below the threshold for decay into fermion pairs. It immediately follows from (\ref{Poles2D}) that $\Omega_1 \le 2\Delta_\Tr{c}$ and $|\Omega_2| \le 2\Delta_\Tr{v}$ at $q=0$ for any $\beta$. As a consequence, at any finite $\beta$ the superfluid transition $\Omega=0$ occurs when $\Delta_\Tr{c}$ and $\Delta_\Tr{v}$ are still both finite. Fermionic excitations are gapped at the transition and may be integrated out from the microscopic action to yield an effective bosonic action which describes the transition. The insulator adjacent to the superfluid is a bosonic Mott insulator.

\subsection{$d=3$}

Following the same procedure as before and in Ref.\cite{nikolic:033608}, we find from (\ref{ppBubble}) the regularized boson propagator $D(q_\mu)=-\pi(q_\mu)^{-1}$:
\begin{eqnarray}\label{ppBubble3D}
\pi(q_\mu) & = & \frac{m\nu}{4\pi} +
    \frac{m_\Tr{c}\gamma_\Tr{c}^2}{4\pi} \sqrt{\frac{q^2}{4}+m_\Tr{c}(2\Delta_\Tr{c}-\Omega)-i0^+} \nonumber \\
&+& \frac{m_\Tr{v}\gamma_\Tr{v}^2}{4\pi} \sqrt{\frac{q^2}{4}+m_\Tr{v}(2\Delta_\Tr{v}+\Omega)+i0^+} \ .
\end{eqnarray}
Again, there are branch-cut singularities for real frequencies $\Omega>\omega_\Tr{c}$ and $\Omega<-\omega_\Tr{v}$ defined in (\ref{omegaPrm}). Let us define mass scales $m$ and $\delta m$ by:
\begin{equation}
m_\Tr{c}^{\frac{3}{2}}\gamma_\Tr{c}^2 = m^{\frac{3}{2}}+\delta m^{\frac{3}{2}} \quad , \quad
m_\Tr{v}^{\frac{3}{2}}\gamma_\Tr{v}^2 = m^{\frac{3}{2}}-\delta m^{\frac{3}{2}} \ ,
\end{equation}
allowing $\delta m^{3/2}$ to be either positive or negative, and then write
\begin{eqnarray}\label{ppBubble3Db}
\!\!\!\!\! \pi(q_\mu) \!\! & = & \!\! \frac{m\Wt{\nu}}{4\pi} + \frac{m}{4\pi} \left(
  \sqrt{m(\omega_\Tr{c}-\Omega)} + \sqrt{m(\omega_\Tr{v}+\Omega)} \right) \\
\!\!\!\!\! \Wt{\nu} \!\! & = & \!\! \nu + \left( \frac{\delta m}{m} \right)^{\frac{3}{2}} \left(
  \sqrt{m(\omega_\Tr{c}-\Omega)} - \sqrt{m(\omega_\Tr{v}+\Omega)} \right) \ . \nonumber
\end{eqnarray}
The solutions of $\pi(q_\mu)=0$ define bosonic poles of either particle $\Tr{Re} \lbrace \Omega \rbrace > 0$ or hole $\Tr{Re} \lbrace \Omega \rbrace < 0$ flavor. The conditions for the existence of the particle-like and hole-like poles at a wavevector $q$ are $\nu<\nu_\Tr{c}(q)$ and $\nu<\nu_\Tr{v}(q)$ respectively, where
\begin{eqnarray}\label{Nu0}
\nu_\Tr{c}(q) \!\! & = & \!\! - \left( 1 - \left( \frac{\delta m}{m} \right)^{\frac{3}{2}} \right)
  \sqrt{m(\omega_\Tr{c}+\omega_\Tr{v})} \\
\nu_\Tr{v}(q) \!\! & = & \!\! - \left( 1 + \left( \frac{\delta m}{m} \right)^{\frac{3}{2}} \right)
  \sqrt{m(\omega_\Tr{c}+\omega_\Tr{v})} \nonumber
\end{eqnarray}
following from the requirement that the square roots in (\ref{ppBubble3Db}) have positive real parts. An example is shown in Fig.\ref{PiSol}(b). The detuning parameter values $\nu=\nu_\Tr{c}(q)$ and $\nu=\nu_\Tr{v}(q)$ correspond to resonant scattering between fermion particles and holes respectively, since they separate the regimes in which two-particle or two-hole bound states exist (BEC) or do not exist (BCS). When the poles exist they can be determined by solving
\begin{equation}\label{Poles3D}
\Omega_{1/2} = \frac{\omega_\Tr{c}-\omega_\Tr{v}}{2} \pm
  \sqrt{\left(\frac{\omega_\Tr{c}+\omega_\Tr{v}}{2}\right)^2
    -\left(\frac{\omega_\Tr{c}+\omega_\Tr{v}}{2}-\frac{\Wt{\nu}^2}{2m}\right)^2}
\end{equation}
whose structure is the same as in (\ref{Poles2D}), allowing us to immediately read out the condition (\ref{PhTrans2D}) for the phase transition (set $\Omega=0$ and $q=0$),
\begin{equation}\label{PhTrans3D}
4 \Delta_\Tr{c} \Delta_\Tr{v} = \left( \Delta_\Tr{c}+\Delta_\Tr{v} - \frac{\Wt{\nu}^2}{2m} \right)^2 \ .
\end{equation}
The transition becomes pair-breaking if it occurs at a vanishing fermion gap $\Delta_\Tr{c}=0$ or $\Delta_\Tr{v}=0$. The former requires $\Wt{\nu}=-\sqrt{2m\Delta_\Tr{v}}$ or equivalently $\nu=\nu_\Tr{c}(0)$, and the latter requires $\Wt{\nu}=-\sqrt{2m\Delta_\Tr{c}}$ or $\nu=\nu_\Tr{v}(0)$.

In the BCS regimes there are no bosonic poles, so that the only path to superfluidity involves the standard BCS instability when a fermion gap is closed to produce a finite quasiparticle density in the ground state. In the BEC regimes, real poles exist in the insulating state (the occurrence of complex poles in (\ref{Poles3D}) is preempted by a superfluid transition). Both the quasiparticle and quasihole gaps, $\Delta_\Tr{c}$ and $\Delta_\Tr{v}$ respectively are finite at the transition in the BEC regimes, so that this transition is in a bosonic universality class.

\section{Discussion and conclusions}

Attractive interactions between fermionic quasiparticles which remain gapped across the superconducting transition give rise to a protected bosonic Mott insulator of Cooper pairs. The superconducting transition at a fixed density is associated with phase fluctuations of the order parameter, and the insulating state adjacent to the superconductor is characterized by gapped low-energy collective modes with bosonic statistic. In two dimensions this Cooper pair insulator completely surrounds the superconducting phase in the phase diagram, making a direct transition between the superconductor and an unpaired band insulator impossible. Above two dimensions, the bosonic Mott physics belongs to the BEC regime of sufficiently strong coupling, while the transition is pair-breaking in the weak coupling limit. Although sufficiently strong pairing and perturbations can stabilize insulating symmetry broken phases of Cooper pairs, there is no a priori reason for this Mott insulator to be thermodynamically distinguished from the band insulator. A sharp distinction nevertheless exists in non-equilibrium settings, and can be characterized as a non-equilibrium phase transition.

The most abstract and unbiased way of identifying transitions must be representation independent. Such information is contained only in the energy spectrum of the Hamiltonian, or density matrix as a generalization which can describe coupling to the environment. Non-analytic changes of the total density of states in the ground state manifold are associated with thermodynamic quantum phase transitions. Analogous non-analytic changes in the excited state spectrum are possible, and can show up as sharp phase-transition-like phenomena in systems driven out of equilibrium. We demonstrated that the presence of Cooper pairs in the bosonic Mott insulator can at least be identified with such a non-equilibrium transition, and physically associated with the ability to drive Bose condensation among excited states into a Cooper pair laser, the state analogous to the many-body state of photons produced by a laser.

We considered a band insulator subjected to pairing as a model system. The simplest realization of such a system is found in trapped neutral ultra-cold gases of alkali atoms placed in an optical lattice. The density of atoms can be chosen to correspond to two atoms per lattice site in the central portion of the trap, while the strength of attractive interactions among them is routinely controlled by the Feshbach resonance. A superfluid transition from a thermally excited band insulator has been already experimentally studied in this kind of a system in the vicinity of the BCS-BEC crossover \cite{Chin2006}. Detecting a Bose-insulator behavior may be more difficult, as it requires very low temperatures and a way to sharply distinguish the presence of Cooper pairs (locally stable pairs have been observed  \cite{Partridge2005, Schunck07}). Perhaps the excited-state condensation might be possible to drive by an external low-frequency radiation, and then detect in the momentum distribution of particles in a time-of-flight measurement.

In condensed matter systems, a two-dimensional band insulator could be rendered unstable to Cooper pairing by the proximity effect to a superconducting layer (in a manner similar to Ref.\cite{Kasumov1996}). The challenge here is to find a material whose bandgap is comparable to the typical critical temperature of the superconducting layer. Bismuth antimony alloys seem to be promising candidates, since their bandgap can be tuned by the antimony concentration between zero and at least 25 meV \cite{Golin1968}. Another solution might be to exploit a charge-density-wave or an engineered superlattice to create a small bandgap. Detecting bosonic excitations in such insulators also seems difficult, but indirect indications might be observable in transport measurements, such as Nernst effect in the vicinity of the superconducting transition. The Cooper pair insulator physics might also be relevant for the amorphous oxide films.

Perhaps the most promising way to test the Cooper pair insulator hypothesis is by Monte-Carlo simulations in negative-$U$ Hubbard models \cite{Randeria1992}. External driving mechanism can be simulated, and it should be possible to observe the formation of a driven condensate, and the corresponding non-equilibrium transition to a true band insulator.

\subsection{A qualitative picture of cuprates}

The applicability of the picture presented in this paper to cuprates depends on two things: (i) whether fermionic excitations acquire at least a partial gap independent of pairing \cite{Hufner2008}, for example due to spin correlations or fluctuations
in the particle-hole channel, and (ii) whether we can understand
the two-dimensional $d$-wave superconductivity in the same spirit as
a BEC regime despite some obvious differences, like
the absence of electron pair bound states.

Phenomenologically, an independent antinodal gap indeed exists and is reflected by the $T^*$ crossover temperature in underdoped cuprates which grows the closer one gets to the antiferromagnetic (AF) phase. In the context of $t-J$ models one could naively relate the fermion gap to the exchange energy cost $\sim J \langle S \rangle$ for removing a spin from the local AF environment, where $\langle S \rangle$ is the effective
local magnetic moment per site, reduced by quantum fluctuations due to doping. There are a number of sophisticated theories dealing with the microscopic origin of such a gap \cite{anderson87c, Wen1995, senthil00, Chakravarty2001, anderson04, Kaul2007}. For our purposes, it is sufficient to assume that the gap originates from spin dynamics and that it decreases with doping. This would ensure an antinodal gap dominated by the nearest-neighbor spin exchange.

Next, we recall an old idea that charge organized in the form of Cooper pairs
could propagate through a short-range correlated AF environment without frustration.
A familiar caricature is that of a hole moving by nearest-neighbor hopping
and leaving a trail of frustrated spins along its path; a second hole which follows right behind would repair this damage and thus the two are
"attracted" to each other by the intervention of the AF environment
in which both move. Some variation of this mechanism -- whose
precise form is unknown at present time -- then suffices
to produce an {\em effective} attractive short-range interaction
between holes, which unavoidably has anisotropic properties,
being the strongest along the lattice bond directions.
Certainly, this interaction has to fight against ``quantum entropy''
and, at a given low doping, need not be strong enough to actually cause pairing,
let alone superconductivity. However, its anisotropy
ensures that antinodal holes are most affected.

Therefore, we arrive at a situation which is similar to the one
analyzed in this paper. Fermionic charged excitations near
antinodal regions of the Brillouin zone are gapped, but
experience short-range (anisotropic) pairing  interactions.
If the concentration of holes is increased, the gap decreases, while the effect of interactions becomes stronger because the average separation between holes shrinks. These trends persist until the characteristic interaction scale becomes comparable with the gap scale, and then the conditions for superconducting instability are created. The analysis in this paper shows that bosonic excitations with charge $2e$ and no spin necessarily play a crucial role in these circumstances. The short-range ``antinodal'' Cooper pairs have a certain robustness across the superconducting transition and provide a backbone for a correlated ``normal'' state.

However, the $d$-wave order parameter symmetry requires that nodal quasiparticles remain gapless across the transition,\cite{tesanovic04}
which makes cuprates fundamentally different from the models studied in this paper. A conventional BEC regime is out of question since it requires a full quasiparticle gap and a Cooper pair bound state. It is useful
to think of the difference between the two in the real space. Deep in the conventional BEC
regime, tightly bound Cooper pairs have a small typical size, smaller than the
average separation between particles, and the pairs much
larger than this size are absent for all practical purposes. In the nodal $d$-wave
case there is a {\em power-law distribution} of the pair sizes and thus
there are {\em arbitrarily large} Cooper pairs, testifying to the presence of nodes.
Nevertheless, we expect that in cuprates most of the Cooper pairs are still
rather small, their size basically given by the Cu-O-Cu bond length. It is these
small, stick-like Cooper pairs that, within our picture of the pairing gap,
are responsible for the antinodal gap. These bond Cooper pairs
can actually be represented by
an effective bosonic theory of the type discussed in this paper.\cite{tesanovic04}
While such purely bosonic theory is incomplete when applied to cuprates and
must be supplemented by the nodal fermionic quasiparticles, a subset of its properties
in the {\em charge} sector is in fact shared by the tightly bound BEC-limit pairs. For
example, there will be preference for certain ``magic dopings" where Cooper
pairs can neatly fit into the underlying lattice, although the resulting
``pair density wave" patterns will in general be rather different -- the reason is
that, even at this level, there is a difference from $s$-wave description since
the stick-like nature of $d$-wave Cooper pairs in cuprates makes them
unusually prone to formation of various quasi one-dimensional
and nematic patterns ordinarily absent
from the conventional BEC limit.\cite{tesanovic04}
Very importantly, the decay channel for Cooper pairs in
cuprates is anomalously slow due to the linearly
vanishing density of quasiparticle states at the
gapless points in the Brillouin zone. The resulting longevity of Cooper pairs is what
likely protects the integrity of bosonic fluctuations in the pseudogap state,
both in terms of pairbreaking and vortex dynamics.\cite{footnote}

Thus, we hypothesize that charged quasiparticles in cuprates
live in an {\em unconventional} BEC-like regime due to both quasi-two-dimensionality and relatively large strength of interactions. This BEC regime is peculiar since it does not involve low-energy bound-state pairs as a result of the $d$-wave pairing symmetry.
Consequently, it lacks a true, sharp charge gap and it clearly has gapless nodal
spin excitations. While some of the features of this picture can be
gleamed from the d-wave dual approach, \cite{Tesanovic2008}
many details needed to fully verify its validity are yet to be worked out. For example, the emergence of this peculiar BEC regime from within a renormalization group analysis \cite{Nikolic2010} might require non-analytic features of the effective interactions between quasiparticles at zero momentum transfer. In any event, this may be the route to ultimately understanding a variety of unconventional properties of cuprates as manifestations of quantum vortex dynamics, including the Nernst effect and density-wave patterns observed by Scanning Tunneling Microscopy \cite{tesanovic04, balents05, nikolic:134511, nikolic:144516}.

At very low dopings, close to the AF phase, the dominant low-energy excitations are bosons which carry spin $S=1$, but no charge. Therefore, a certain crossover between spin-dominated and charge-dominated regime is expected in the pseudogap state at zero temperature. It is natural to interpret the region in the cuprate phase diagram where the Nernst signal is large as a charge-dominated regime, below a dome-shaped onset temperature for pairing which encloses the superconducting dome in the phase diagram of cuprates.


\section{Acknowledgements}
We are grateful to E. Zhao and V. Stanev for useful discussions. The support for this research was provided by the Office of Naval Research (grant N00014-09-1-1025A), and the National Institute of Standards and Technology (grant 70NANB7H6138, Am 001). Work at the Johns Hopkins-Princeton Institute for Quantum Matter was supported by the U.\ S.\ Department of Energy, Office of Basic Energy Sciences, Division of Materials Sciences and Engineering, under Award No.\ DE-FG02-08ER46544.

\appendix

\section{Effective theories of pairing in band insulators}\label{appBoseModels}

Here we derive effective low-energy bosonic lattice theories from the microscopic continuum model (\ref{ContModel1}) in the regime where the lowest energy excitations with bosonic statistics lie well below twice the finite gap for fermionic excitations. We can introduce a complex Hubbard-Stratonovich field $\Phi(\rv_1,\rv_2)$ to decouple the interaction term in  (\ref{ContModel1}):
\begin{eqnarray}\label{Cont2Ch1}
&& S = \int \dd\tau \biggl\lbrace \dd^{d}r \psi_{\alpha}^{\dagger}
  \left( \frac{\partial}{\partial\tau} - \frac{\nabla^2}{2m} + V(\boldsymbol{r}) - \mu \right) \psi_{\alpha} \nonumber \\
&& ~~ + \int \dd^{d}r_1 \dd^{d}r_2 \biggl\lbrack
     \frac{1}{U(|\rv_1-\rv_2|)} \Phi(\rv_1,\rv_2)^{\dagger} \Phi(\rv_1,\rv_2)^{\phantom{\dagger}} \nonumber \\
&& ~~ + \Phi(\rv_1,\rv_2)^{\phantom{\dagger}} \psi_{\uparrow}^{\dagger}(\rv_1) \psi_{\downarrow}^{\dagger}(\rv_2) +
     h.c.
     \biggr\rbrack \biggr\rbrace
\end{eqnarray}
In order to rewrite the model in a tight-binding form, we represent the field operators as expansions weighted by Wannier functions:
\begin{eqnarray}
\psi_\alpha(\rv) & = & \sum_{n,\Rv} W_{n\Rv}(\rv) f_{\alpha n\Rv} \\
\Phi(\rv_1,\rv_2) & = & \sum_{n_1,\Rv_1} \sum_{n_2,\Rv_2} \widetilde{W}_{n_1\Rv_1}(\rv_1) \widetilde{W}_{n_2\Rv_2}(\rv_2)
  b_{n_1\Rv_1,n_2\Rv_2} \nonumber
\end{eqnarray}
Here, $\Rv$ are the discrete coordinates of lattice sites, $n$ band-index or orbital labels, $f_{\alpha n\Rv}$ fermionic lattice fields and $b_{n_1\Rv_1,n_2\Rv_2}$ bosonic lattice fields. The bosonic Wannier functions $\widetilde{W}$ can be chosen at will (subject to constraints discussed below), while the best choice for the fermionic Wannier functions $W$ is dictated by the bare fermion dispersion in the lattice potential. If $\psi_{n\kv}$ are Bloch wavefunctions in the potential $V(\rv)$, then the choice
\begin{equation}
W_{n\Rv}(\rv) = \int_{B.Z.} \frac{\dd^d k}{(2\pi)^d} e^{-i\kv\Rv} \psi_{n\kv}(\rv)
\end{equation}
allows us to express the fermionic kinetic part of the action as
\begin{eqnarray}
S_f & = & \int \dd\tau \Biggl\lbrack \sum_{n\Rv}
    f_{\alpha n\Rv}^{\dagger} \left( \frac{\partial}{\partial\tau} - \mu \right) f_{\alpha n\Rv}^{\phantom{\dagger}} \\
 & - & \sum_n \sum_{\Rv_1\Rv_2} t_{n,\Rv_1-\Rv_2}^{\phantom{\dagger}}
       f_{\alpha n\Rv_1}^{\dagger} f_{\alpha n\Rv_2}^{\phantom{\dagger}} \Biggr\rbrack \nonumber \ ,
\end{eqnarray}
where the hopping couplings are functions of the bare energies $\epsilon_{n\kv}$:
\begin{equation}
t_{n\Bf{\Delta R}} = - \int_{B.Z.} \frac{\dd^d k}{(2\pi)^d} e^{-i\kv\Bf{\Delta R}} \epsilon_{n\kv} \ .
\end{equation}
The quadratic bosonic term becomes:
\begin{eqnarray}
S_b & = & \int \dd\tau \sum_{n_1\Rv_1} \sum_{n_2\Rv_2} \sum_{n'_1\Rv'_1} \sum_{n'_2\Rv'_2}
  K_{n_1\Rv_1,n'_1\Rv'_1,n_2\Rv_2,n'_2\Rv'_2} \times \nonumber \\
  & & b_{n_1\Rv_1,n_2\Rv_2}^{\dagger} b_{n'_1\Rv'_1,n'_2\Rv'_2}^{\phantom{\dagger}} \ ,
\end{eqnarray}
where
\begin{eqnarray}
&& K_{n_1\Rv_1,n'_1\Rv'_1,n_2\Rv_2,n'_2\Rv'_2} = \int \dd^d r_1 \dd^d r_2
  \frac{1}{U(|\rv_1-\rv_2|)} \times  \nonumber \\
&& ~~~ \widetilde{W}_{n_1\Rv_1}^*(\rv_1) \widetilde{W}_{n'_1\Rv'_1}^{\phantom{*}}(\rv_1)
        \widetilde{W}_{n_2\Rv_2}^*(\rv_2) \widetilde{W}_{n'_2\Rv'_2}^{\phantom{*}}(\rv_2) \ .
\end{eqnarray}
The localized nature of Wannier functions will generally result with significant values for $K$ only when $\Rv_1\approx\Rv'_1$ and $\Rv_2\approx\Rv'_2$. On the other hand, the interaction potential $U(|\rv_1-\rv_2|)$ controls the dependence of $K$ on $\Bf{\Delta}\Rv=\Rv_1-\Rv_2$ and $\Bf{\Delta}\Rv'=\Rv'_1-\Rv'_2$, causing $K$ to become larger and larger as $\Bf{\Delta}\Rv$ and $\Bf{\Delta}\Rv'$ grow. Since $K$ acts like a mass term for the $b$ fields, fluctuations of long-range singlet bonds will be suppressed. The requirement that $K$ be convergent puts certain restrictions on the choice of $\widetilde{W}$. Namely, any short-range interaction potential $U(|\rv_1-\rv_2|)$ will generate diverging factors in the limit $|\rv_1-\rv_2|\to\infty$ which must be controlled and tamed by sufficiently fast decaying Wannier functions. With the exception of the contact potential $U \propto \delta(\rv_1-\rv_2)$ which we will separately discuss later, we shall assume for simplicity that $U(|\rv_1-\rv_2|)$ is not too short ranged, so that we can use the same Wannier functions for the boson and fermion fields, $\widetilde{W}_{n\Rv}(\rv)=W_{n\Rv}(\rv)$. Then, the interaction term takes a simple form:
\begin{equation}
S_{bf} = \int \dd\tau \sum_{n_1\Rv_1} \sum_{n_2\Rv_2} \Bigl(
  b_{n_1\Rv_1,n_2\Rv_2}^{\phantom{\dagger}} f_{\uparrow n_1\Rv_1}^{\dagger} f_{\downarrow n_2\Rv_2}^{\dagger}
  + h.c. \Bigr) \ .
\end{equation}

The full lattice action $S=S_f+S_b+S_{bf}$ is an exact rewriting of the starting model (\ref{ContModel1}). We can introduce a short-hand notation for $(n,\Rv)$ by defining a ``hyper-lattice'' whose sites $i,j\dots$ carry both the regular lattice coordinates $\Rv$ and band/orbital coordinates $n$:
\begin{eqnarray}\label{LatticeModel1}
S & = & \int \dd\tau \biggl\lbrack \sum_i
        f_{i\alpha}^{\dagger} \left( \frac{\partial}{\partial\tau} - \mu \right) f_{i\alpha}^{\phantom{\dagger}}
        -\sum_{ij} t_{ij}^{\phantom{\dagger}} f_{i\alpha}^{\dagger} f_{j\alpha}^{\phantom{\dagger}} \nonumber \\
  & + & \sum_{ijkl} K_{ijkl}^{\phantom{\dagger}} b_{ij}^{\dagger} b_{kl}^{\phantom{\dagger}}
      + \sum_{ij} \left( b_{ij}^{\phantom{\dagger}} f_{i\uparrow}^{\dagger} f_{j\downarrow}^{\dagger} + h.c. \right)
  \biggr\rbrack \ .
\end{eqnarray}

Since we assume that the chemical potential $\mu$ lies in a bandgap, the bare fermionic fields are fully gapped and safe to integrate out. We are interested in the low-energy dynamics of singlet bonds or Cooper pairs $b_{ij}$, so we also integrate out all high energy $b_{ij}$ fields obtained for too large spatial separations between sites $i$ and $j$. The effective theory containing only the degrees of freedom below the fermion bandgap scale is a purely bosonic lattice model in which the $b$ fields acquire effective dynamics. We cannot derive the exact form of the effective action, but schematically:
\begin{eqnarray}\label{EffModel1}
S_{\Tr{eff}}^{(1)} & = & \int \dd\tau \biggl\lbrace \sum_{ijkl} b_{ij}^{\dagger}
    K_{ijkl}^{\Tr{eff}}\left[\frac{\partial}{\partial\tau}\right] b_{kl}^{\phantom{\dagger}} \\
  & + & \sum_{ijkl} \sum_{i'j'k'l'} \mathcal{U}_{ijkl}^{i'j'k'l'} b_{ij}^{\dagger} b_{kl}^{\dagger}
    b_{i'j'}^{\phantom{\dagger}} b_{k'l'}^{\phantom{\dagger}} + \cdots \biggr\rbrace \ . \nonumber
\end{eqnarray}
Since only the short-range bonds $b_{ij}$ remain, the number of bosonic degrees of freedom scales as the number of lattice sites, with an addition of a finite number of low-energy ``orbital'' states per spatial lattice site. Without knowing the couplings, this action is very general. We can be only certain that it captures a superfluid-insulator transition, because the ``two-channel'' model (\ref{Cont2Ch1}) is guarantied to prefer the condensation of $\Phi$ at any $U^{-1}$ when $\mu$ approaches close enough to a band-edge. Since ``orbital'' states are separated by finite gaps, only one bosonic mode becomes soft at the superfluid transition, implying that a different bosonic effective model, with one degree of freedom per site, is also sufficient to describe the universal properties of this transition. Below we outline the derivation of such an effective model, which discards all microscopic details of the interaction potential $U(|\Bf{\Delta}\rv|)$ and replaces it by a contact interaction. However, keeping these microscopic details in the action may be needed for the description of other transitions between the insulating states. For example, should in some circumstances the singlet bonds become strongly repulsive, yet close-packed, the effective theory of this kind would be able to describe a variety of Mott insulating states, including valence-bond states.

If one is mainly interested in universal properties, one identifies a renormalization group fixed point and considers only the scale dependence of relevant operators. The microscopic details of the interaction potential $U$ are found to be irrelevant at least at the Gaussian fixed point, and at unitarity $U\to\infty$ in (\ref{Cont2Ch1}) in dimensions $d \ge 2$. The only relevant piece is the contact interaction $U(|\Bf{\Delta}\rv|) \to U\delta(\Bf{\Delta}\rv)$, for which the above bond effective theory breaks down. Instead, one may decouple the interaction by a Hubbard-Stratonovich field $\Phi(\rv)$, and define on-site bosonic fields $b$ by:
\begin{equation}
\Phi(\rv) = \sum_{n,\Rv} \widetilde{W}_{n\Rv}(\rv) b_{n\Rv}
\end{equation}
The full lattice theory takes the form:
\begin{eqnarray}\label{LatticeModel2}
\!\!\!\!\!\!\!\!\!\! S & = & \int \dd\tau \biggl\lbrack \sum_i
        f_{i\alpha}^{\dagger} \left( \frac{\partial}{\partial\tau} - \mu \right) f_{i\alpha}^{\phantom{\dagger}}
        -\sum_{ij} t_{ij}^{\phantom{\dagger}} f_{i\alpha}^{\dagger} f_{j\alpha}^{\phantom{\dagger}} \nonumber \\
\!\!\!\!\!\!\!\!\!\! & + & U^{-1} \sum_{i} b_i^{\dagger} b_i^{\phantom{\dagger}}
      + \sum_{ijk} g_{ijk} \left( b_i^{\phantom{\dagger}} f_{j\uparrow}^{\dagger} f_{k\downarrow}^{\dagger} + h.c. \right)
  \biggr\rbrack \ , ~~
\end{eqnarray}
and the effective pure bosonic theory is:
\begin{equation}\label{EffModel2}
S_{\Tr{eff}}^{(2)} = \int \dd\tau \biggl\lbrace \sum_{ij} b_i^{\dagger}
    K_{ij}^{\Tr{eff}}\left[\frac{\partial}{\partial\tau}\right] b_j^{\phantom{\dagger}}
  + \sum_{ijkl} \mathcal{U}_{ij}^{kl} b_i^{\dagger} b_j^{\dagger}
    b_k^{\phantom{\dagger}} b_l^{\phantom{\dagger}} + \cdots \biggr\rbrace \ .
\end{equation}

\section{The range of effective interactions between bosonic excitations in the Mott insulator}\label{appInt}

Here we outline the calculation of the diagram in Fig.\ref{BosonInteraction}(b) which describes the scattering processes of charged bosonic excitations. We assume that the ground state is a bosonic Mott insulator so that coherent bosonic excitations exist and all excitations are gapped. We are interested in low energy bosons which cannot decay into fermions during the course of a collision. The boson and fermion gaps are $\Delta_\Tr{b}$ and $\Delta_\Tr{f}$ respectively, where $\Delta_\Tr{b} < 2\Delta_\Tr{f}$.

Following Feynman rules and writing $k_\mu=(\omega,\kv)$, $q_{i\mu}=(\Omega_i,\qv_i)$, $p_\mu=(\Omega,\pv)$ we start from the generic scattering diagram in Fig.\ref{BosonInteraction}(b):
\begin{widetext}
\begin{eqnarray}
&& \!\!\!\!\!\!\! U_\Tr{b}(q_{1\mu},q_{2\mu};p_\mu) = i \int \frac{\dd\omega}{2\pi} \frac{\dd^d k}{(2\pi)^d} \;
  G\left(k_\mu+\frac{p_\mu}{2}\right) G\left(k_\mu-\frac{p_\mu}{2}\right)
  G\left(-k_\mu+q_{1\mu}-\frac{p_\mu}{2}\right) G\left(-k_\mu+q_{2\mu}+\frac{p_\mu}{2}\right) \\
&& = i \int \frac{\dd\omega}{2\pi} \frac{\dd^d k}{(2\pi)^d} \;
  \frac{1}{\omega+\frac{\Omega}{2}-\EE_{\kv+\frac{\pv}{2}}} \;
  \frac{1}{\omega-\frac{\Omega}{2}-\EE_{\kv-\frac{\pv}{2}}} \;
  \frac{1}{-\omega+\Omega_1-\frac{\Omega}{2}-\EE_{-\kv+\qv_1-\frac{\pv}{2}}} \;
  \frac{1}{-\omega+\Omega_2+\frac{\Omega}{2}-\EE_{-\kv+\qv_2+\frac{\pv}{2}}} \nonumber \\
&& = \int \frac{\dd^d k}{(2\pi)^d} \; \frac{1}{\alpha_2-\alpha_1} \left\lbrack
  \frac{1}{\alpha_1-\EE_{\kv+\frac{\pv}{2}}+\frac{\Omega}{2}} \;
  \frac{1}{\alpha_1-\EE_{\kv-\frac{\pv}{2}}-\frac{\Omega}{2}} -
  \frac{1}{\alpha_2-\EE_{\kv+\frac{\pv}{2}}+\frac{\Omega}{2}} \;
  \frac{1}{\alpha_2-\EE_{\kv-\frac{\pv}{2}}-\frac{\Omega}{2}} \right\rbrack \nonumber \\
&& = \int \frac{\dd^d k}{(2\pi)^d} \; \frac{\alpha_1+\alpha_2-\EE_{\kv+\frac{\pv}{2}}-\EE_{\kv-\frac{\pv}{2}}}
  {\left(\alpha_1-\EE_{\kv+\frac{\pv}{2}}+\frac{\Omega}{2}\right)
   \left(\alpha_1-\EE_{\kv-\frac{\pv}{2}}-\frac{\Omega}{2}\right)
   \left(\alpha_2-\EE_{\kv+\frac{\pv}{2}}+\frac{\Omega}{2}\right)
   \left(\alpha_2-\EE_{\kv-\frac{\pv}{2}}-\frac{\Omega}{2}\right)}
  \nonumber
\end{eqnarray}
\end{widetext}
Assuming that all fermionic excitations are particle-like, with energy
\begin{equation}
\EE_{\kv} = \Delta_\Tr{f} + \frac{k^2}{2m} > 0 \ ,
\end{equation}
each denominator in the second line contains the same implicit infinitesimal term $i0^+$. The internal frequency integral extended to a loop over the upper complex half-plane picks the poles of the last two Green's functions. We have also introduced
\begin{eqnarray}
\alpha_1 & = & \Omega_1 - \EE_{-\kv+\qv_1-\frac{\pv}{2}} - \frac{\Omega}{2} \\
\alpha_2 & = & \Omega_2 - \EE_{-\kv+\qv_2+\frac{\pv}{2}} + \frac{\Omega}{2} \ . \nonumber
\end{eqnarray}
for notational brevity, and we will use $\EE_{-\kv}=\EE_{\kv}$ when convenient.

Next, we concentrate on zero energy $\Omega=0$ and finite momentum $p=|\pv|\neq 0$ transfer. This will allow extracting the range of static (non-retarded) effective interactions. Also, let us consider only the lowest energy incoming particles whose momenta are zero, $\qv_1=\qv_2=0$. In order to keep these incoming particles on the mass shell we must also choose $\Omega_1=\Omega_2=\Delta_\Tr{b}$. Then:
\begin{equation}
U_\Tr{b} = \int \frac{\dd^d k}{(2\pi)^d} \; \frac{2 \left(\delta+\frac{k^2}{m}+\frac{p^2}{4m}\right)^{-1}}
       {\left(\delta+\frac{k^2}{m}+\frac{p^2}{4m}\right)^2
         -\left(\frac{\kv\pv}{m}\right)^2 } \ ,
\end{equation}
where $\delta=2\Delta_\Tr{f}-\Delta_\Tr{b}$. It is already clear that $4m\delta$ is the only externally set scale which appears alongside $p^2$. Since the Fourier transform $U_\Tr{b}(p)$ determines the spatial dependence of the effective interaction potential, we see that $r\sim (4m\delta)^{-1/2}$ is a length-scale characterizing this interaction. As long as $\Delta_\Tr{b}<2\Delta_\Tr{f}$, $r$ is finite and the effective interaction is short-ranged.

Further progress can be made in $d=2$ and $d=3$ dimensions. Changing variables to $\kappa=k^2/m$ we obtain in $d=2$:
\begin{eqnarray}
&& \!\!\!\!\!\!\! U^{2D}_\Tr{b} = \frac{m}{(2\pi)^2} \int\limits_0^{2\pi} \dd\theta \int\limits_0^\infty \dd\kappa \;
   \frac{\left(\delta+\frac{p^2}{4m}+\kappa\right)^{-1}}
   {\left(\delta+\frac{p^2}{4m}+\kappa\right)^2-\kappa\frac{p^2}{m}\cos^2\theta} \nonumber \\
&& = \frac{m}{2\pi} \int\limits_0^\infty \dd\kappa \; \frac{\left(\delta+\frac{p^2}{4m}+\kappa\right)^{-2}}
   {\sqrt{\left(\delta+\frac{p^2}{4m}+\kappa\right)^2-\kappa\frac{p^2}{m}}} \\
&& = \frac{2m^3}{\pi} \; \frac{1}{(p^2+4m\delta)^{\frac{3}{2}}} \times \frac{1}{p} \log \left(
   \frac{\sqrt{p^2+4m\delta}+p}{\sqrt{p^2+4m\delta}-p} \right) \nonumber
\end{eqnarray}
This expression is finite for any $p$ as long as $\delta>0$. In $d=3$ we can write $x=\cos\theta$ and immediately integrate out the polar angle:
\begin{eqnarray}
&& \!\!\!\!\!\!\! U^{3D}_\Tr{b} = \frac{m^{\frac{3}{2}}}{(2\pi)^2}
   \int\limits_{-1}^{1} \dd x \int\limits_0^\infty \dd\kappa \;
   \frac{\sqrt{\kappa} \left(\delta+\frac{p^2}{4m}+\kappa\right)^{-1}}
   {\left(\delta+\frac{p^2}{4m}+\kappa\right)^2-\kappa\frac{p^2}{m}x^2} \nonumber \\
&& = \frac{m^2}{2\pi^2} \int\limits_0^\infty \; \frac{\dd\kappa}{\left(\delta+\frac{p^2}{4m}+\kappa\right)^2}
   \times \frac{1}{p} \Tr{arctanh} \left( \frac{p\sqrt{\kappa/m}}{\delta+\frac{p^2}{4m}+\kappa} \right) \nonumber \\
&& \approx \frac{m^3}{2\pi} \; \frac{1}{(p^2+4m\delta)^\frac{3}{2}}
\end{eqnarray}
where in the last step we have taken the limit $p^2 \ll 4m\delta$ before integrating out $\kappa$. Interpreting these expressions as Fourier transforms of the effective interaction potential we clearly find that in both two and three dimensions the effective interaction has a finite range $r\sim(4m\delta)^{-1/2}$.

If incoming particles had finite momenta and energies above the boson gap, no qualitative modifications of these conclusions would be encountered other than $\delta\to2\Delta_f-\EB$, where $\EB$ can be the energy of either incoming boson. This could be interpreted as an energy-dependent effective interaction range, but it also reflects the fact that too energetic particles break into fermion pairs during collisions, so that scattering cannot always be described merely by an interaction potential.

\newpage


\begin{thebibliography}{10}
\bibitem{footnote}
An important special case here is a $d$-wave nodal superconductor
where the decay of Cooper pairs into gapless fermions is always allowed
but anomalously slow due
to the vanishing density of states of nodal quasiparticles
\cite{tesanovic04, nikolic:134511}. In this
situation, both bosonic excitations and nodal fermions must be retained
in the effective theory in the strongly correlated regime,
$\Delta_{\Tr{f}}^{\Tr{max}}\gg \Delta_{\Tr{b}}$, where
$\Delta_{\Tr{f}}^{\Tr{max}}$ is the maximum value of the d-wave gap.
\end{thebibliography}

\end{document}